\documentclass[english,12pt,draftcls,onecolumn]{IEEEtran}
\usepackage[T1]{fontenc}
\usepackage[latin9]{inputenc}
\usepackage{units}
\usepackage{bm}
\usepackage{amsthm}
\usepackage{amsmath}
\usepackage{amssymb}
\usepackage{graphicx}

\makeatletter

\providecommand{\tabularnewline}{\\}

\theoremstyle{plain}
\newtheorem{thm}{\protect\theoremname}
\theoremstyle{plain}
\newtheorem{prop}[thm]{\protect\propositionname}
\theoremstyle{plain}
\newtheorem{lem}[thm]{\protect\lemmaname}

\@ifundefined{showcaptionsetup}{}{%
 \PassOptionsToPackage{caption=false}{subfig}}
\usepackage{subfig}
\makeatother

\usepackage{babel}
\providecommand{\lemmaname}{Lemma}
\providecommand{\propositionname}{Proposition}
\providecommand{\theoremname}{Theorem}

\begin{document}

\title{Tracking the Tracker from its Passive Sonar ML-PDA Estimates}

\author{Domenico~Ciuonzo{*},~\IEEEmembership{Student~Member,~IEEE,} Peter~K.~Willett$\dagger$,~\IEEEmembership{Fellow,~IEEE,}\\
 Yaakov~Bar-Shalom$\dagger$,~\IEEEmembership{Fellow,~IEEE}%
\thanks{P. K. Willett was supported by ONR under contracts N00014-09-10613
and N10014-10-10412. Y. Bar-Shalom was supported by ARO under contract
W911NF-10-10369 and ONR under contract N00014-10-1-0029. %
}\vspace{-1.5cm}
}

\maketitle
\begin{center}
{*} Dept. of Industrial and Information Engineering, Second University
of Naples, Aversa, (CE), Italy. \\
$\dagger$Dept. of Electrical and Computer Engineering, University
of Connecticut, Storrs, (CT), USA.\\
Email: \texttt{domenico.ciuonzo@unina2.it,}~\\
\texttt{ \{willett,ybs\}@engr.uconn.edu}.
\par\end{center}
\begin{abstract}
Target motion analysis with wideband passive sonar has received much
attention. Maximum-likelihood probabilistic data-association (ML-PDA)
represents an asymptotically efficient estimator for deterministic
target motion, and is especially well-suited for low-observable targets;
the results presented here apply to situations with higher signal
to noise ratio as well, including of course the situation of a deterministic
target observed via ``clean'' measurements without false alarms
or missed detections. Here we study the inverse problem, namely, how
to identify the observing platform (following a ``two-leg'' motion
model) from the results of the target estimation process, i.e. the
estimated target state and the Fisher information matrix, quantities
we assume an eavesdropper might intercept. We tackle the problem and
we present observability properties, with supporting simulation results.\end{abstract}
\begin{IEEEkeywords}
Eavesdropper, Fisher information matrix, ML-PDA, nonlinear identification,
platform localization, stealthy platform, wideband passive sonar.
 
\end{IEEEkeywords}

\section{Introduction}

\subsection{Problem Motivation}

\begin{figure}
\subfloat[Collecting measurements.]{\includegraphics[width=0.4\paperwidth]{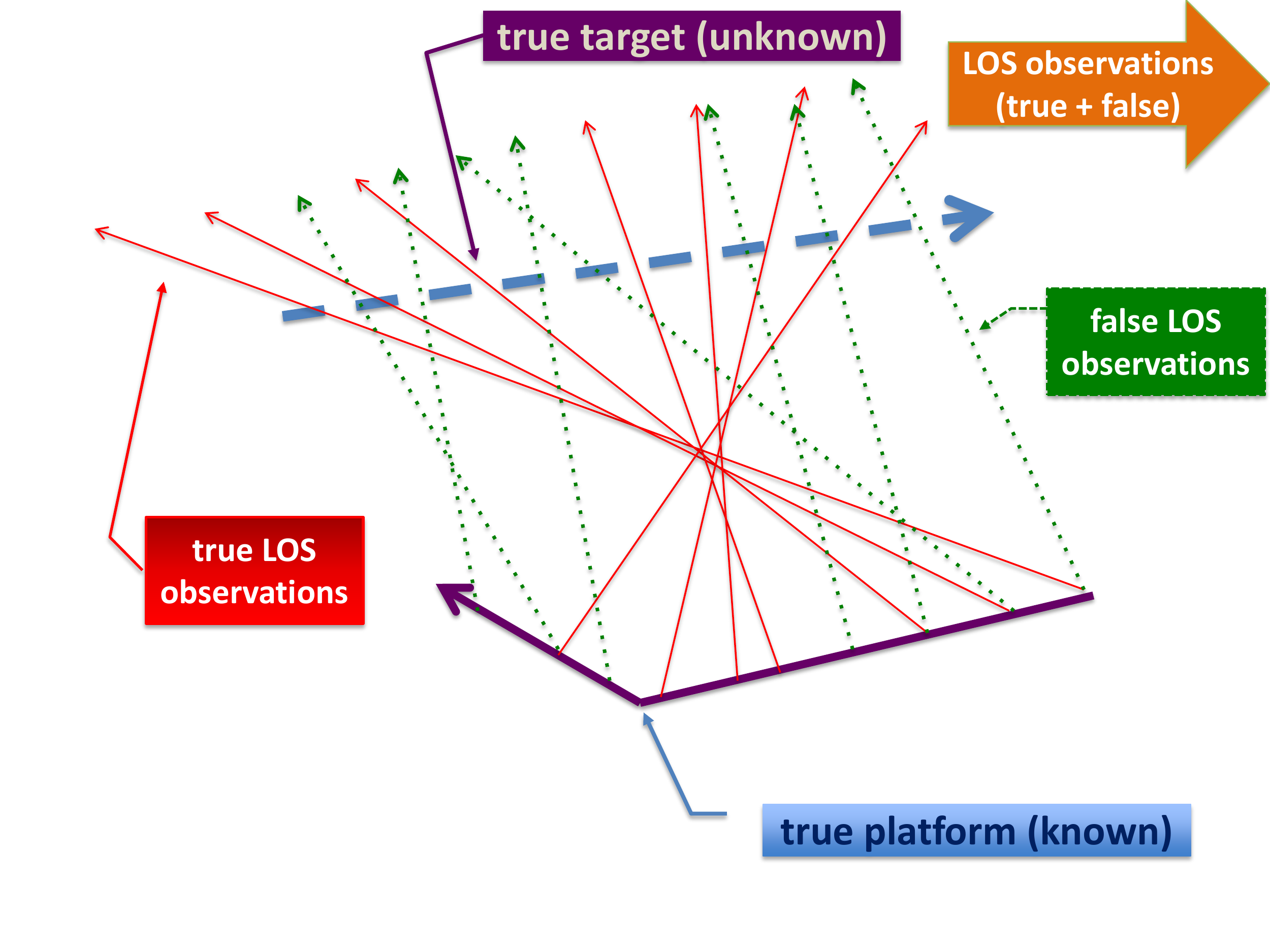}}\subfloat[Estimating target trajectory.]{\includegraphics[width=0.4\paperwidth]{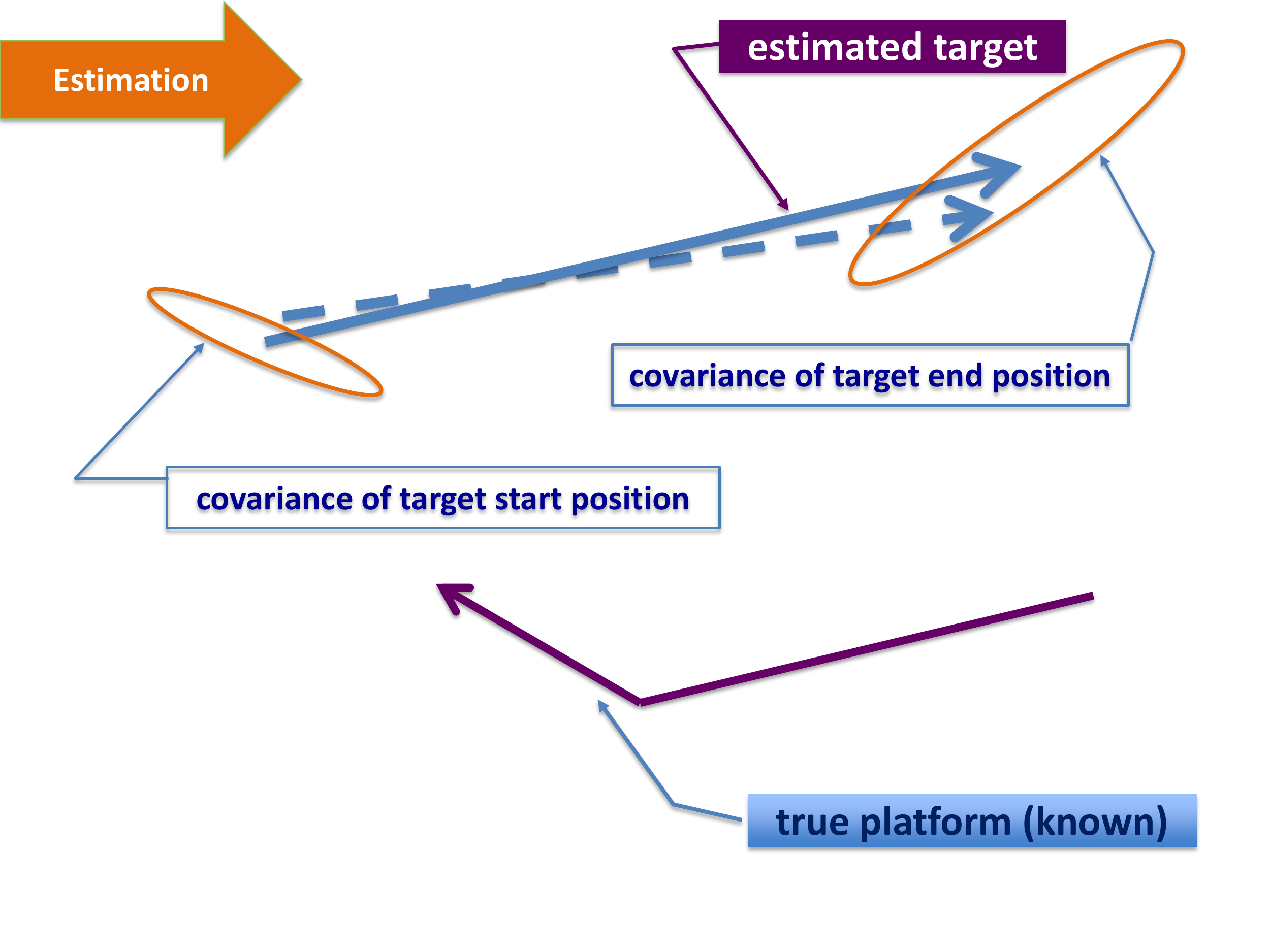}}\caption{The target estimation process by the platform side. \label{fig:Target Estimation}}
\end{figure}

Target motion analysis (TMA) with bearings-only measurements is a
well-understood and extensively studied problem (see Fig. \ref{fig:Target Estimation}).
It has been shown in the literature that as long as the platform is
outmaneuvering the target, observability of the latter is assured
and its motion can be inferred, even from very noisy measurements.
Conversely, it is useful to understand whether, given the results
of TMA estimation, it could be possible to identify, completely or
at least partially, the trajectory of the \emph{observing platform}.

\begin{figure}
\subfloat[Intercepting target estimation results from the platform side.]{\includegraphics[width=0.4\paperwidth]{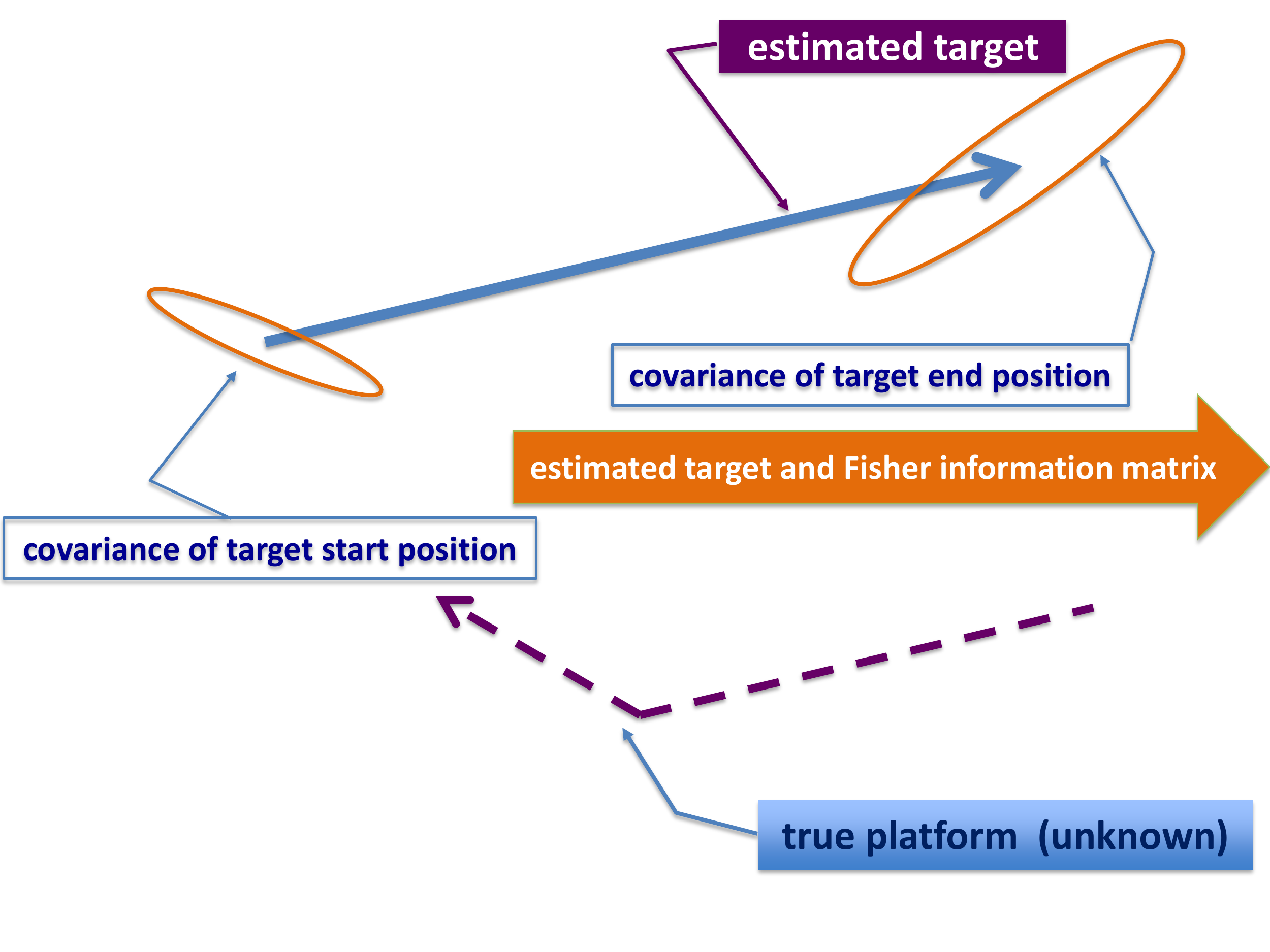}}\subfloat[Identifying the platform trajectory.]{\includegraphics[width=0.4\paperwidth]{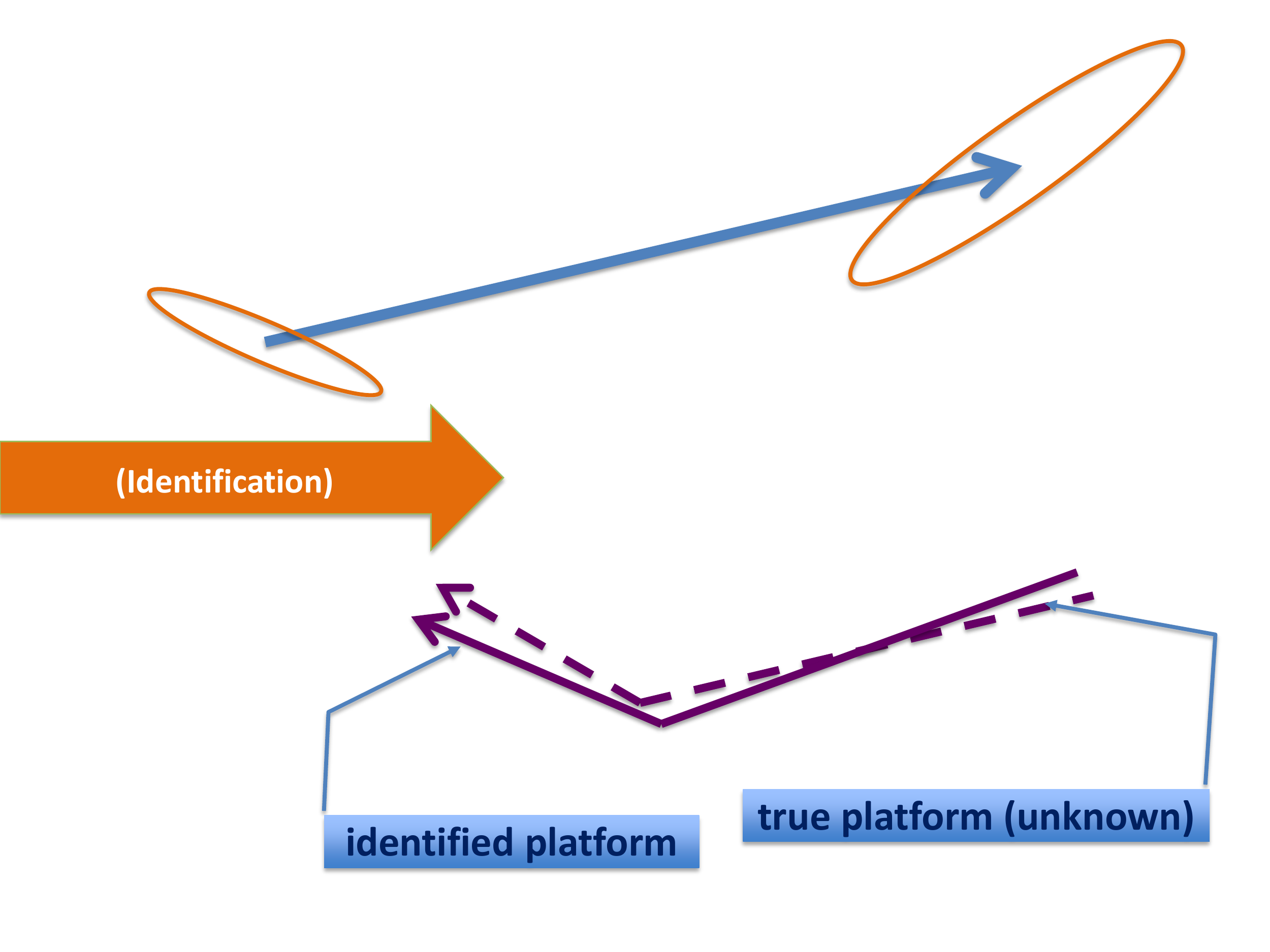}}\caption{The platform identification process by the target-friendly side. \label{fig:Platform identification}}
\end{figure}

More specifically, the problem arises when a ``target-friendly''
entity, as opposed to being cooperative with the platform, intercepts
the results of the target estimation performed by the platform (see
Fig. \ref{fig:Platform identification}); the question is whether
this entity can identify, partially or totally, the trajectory of
the platform. The feasibility of this problem is of twofold interest:
($i$) it verifies the utility of intercepting communications (containing
TMA-related information) between the platform and platform-cooperative
entities, because this information would be useful; and ($ii$) it
motivates, at the platform side, the need for secure and encrypted
transmission of the TMA estimation results.

\subsection{Related Works}

Seminal results on the continuous-time observability of the target
motion, through a wideband passive sonar, were derived in \cite{Nardone1981,Hammel1985}.
In fact, by successive differentiations of the measurement function,
a necessary condition was derived and it was shown that a platform
maneuver is a needed prerequisite to ensure observability of the target;
however unobservable maneuvering-platform trajectories could exist
(i.e. the platform may still take a trajectory wherefrom the target
is unobservable). This analysis was rigorously extended, in the form
of a necessary and sufficient condition, in \cite{Payne1988,Fogel1988,Becker1993,Jauffret1996};
a comprehensive analysis of observability related to practical scenarios
was also conducted in \cite{Jauffret1996}. These results were also
demonstrated in discrete-time in \cite{Cadre1997} via linear algebra;
observability insights in different scenarios were presented, and
also a stochastic observability (and estimability) analysis was performed.
In \cite{Jauffret2007} a theoretical connection between the invertibility
of the Fisher information matrix (FIM) and target (local) observability
was established. In \cite{Passerieux1998} (and references therein)
the optimal platform maneuver was designed, in the sense of the best
estimation accuracy in terms of the FIM.

There are three common approaches to standard TMA (with or without
Doppler measurements): \emph{maximum likelihood} (ML), \emph{pseudolinear}
(PL) and \emph{instrumental variables }(IV) estimation. Although the
first approach is asymptotically efficient, it is complex and therefore
suboptimal solutions are desirable. PL estimation has the advantage
of being in closed form and of easy computation; however it can lead
to severe bias even in favorable conditions \cite{Douganccay2004}.
Consequently IV estimation, yielding estimates with reduced bias,
has seen recent attention \cite{Cadre1999,Dogancay2004,Douganccay2004,Dovganccay2005,Zhang2010}. 

Alternative bearings-only TMA scenarios have been studied recently
in \cite{Bavencoff2006,Jauffret2010,Clavard2011} and ML batch estimators
have been proposed. More specifically, in \cite{Bavencoff2006} the
problem of bearings-only TMA for conditionally-deterministic target
motion and with operational constraints on the platform is tackled
with Markov chain Monte Carlo methods. In \cite{Jauffret2010} TMA
of a maneuvering target and non-maneuvering platform is studied; observability
is established and a batch estimator is proposed. The concept was
later applied to the scenario of a circular constant-speed target
and a non-maneuvering target in \cite{Clavard2011}.

The estimators proposed in these references do not deal with the problem
of false measurements (clutter) and less-than-unity probability of
detection. The seminal work in \cite{Jauffret1990} derived a ML estimate
of target parameters for both wideband and narrowband passive sonars
in the presence of false detections (clutter), based on probabilistic
data association (ML-PDA); the performance of the estimator was evaluated
in terms of the Cramér-Rao lower bound (CRLB). It was shown that the
effect of the clutter on the performance through the CRLB was simply
via a product with a less-than-unity scalar value, called the information
reduction factor (IRF).

The ML-PDA was extended by incorporating amplitude information to
enhance performance in the scenarios of ``low-observable'' (i.e.
low Signal-to-Noise ratio) targets in \cite{Kirubarajan1996}; improved
accuracy and superior global convergence were demonstrated. In \cite{Chummun2002}
ML-PDA was applied to the problem of low-observable target estimation
using electro-optical sensors; also a sliding-window batch approach
for ML-PDA estimation was derived, capable of dealing with temporary
disappearance of targets and/or targets with velocities changing over
time. ML-PDA was also successfully applied to active sonar tracking
in \cite{Blanding2008}, where also an efficient computation of the
ML estimate, namely, directed subspace search (DSS), was derived.
The use of ML-PDA for early track detection with a radar is discussed
in \cite{Bar-Shalom2011}.

\subsection{Main Results and Paper Organization}

The main contributions of the present paper are summarized as follows:
\begin{itemize}
\item We study the ``inverse'' problem of identifying the \emph{platform}
trajectory, following a ``two-leg'' motion model, through its ML-PDA
estimation results on a target; to the best of our knowledge, such
a problem is addressed here for the first time.
\item We derive and study the objective function to be optimized for identifying
the platform trajectory; it is shown that the optimization of this
function depends on neither the IRF nor the measurement variance at
the platform side; that is, the exact%
\footnote{We will show however that for devising an efficient local-optimization
algorithm a range of variability should be given; however the width
of this range does not affect significantly the performance.%
} information to identify the platform trajectory is unnecessary. 
\item Also it is demonstrated that the platform trajectory is \emph{unobservable
}unless it keeps a constant speed during its two different legs.
\item We use an efficient and practical algorithm, based on a \emph{derivative-free}
local search, to solve the nonlinear problem associated with the identification
task. The local optimization routine is initialized from geometric
considerations and exploits the structure of the observed FIM.
\end{itemize}
The paper is organized as follows: In Sec. \ref{sec:System-Model}
we introduce the model for passive wideband-sonar localization and
we give the background on the ML-PDA approach; in Sec. \ref{sec:Inverse-Localization-search}
we formulate the problem of inverse localization and we show some
important identification properties; in Sec. \ref{sec:Initial Guess Choice}
we devise a procedure to compute a good initial estimate as input
for the local optimization routines, while in Sec. \ref{sec:Simulation-Results}
we show, by simulation, the performance of the proposed solution;
some concluding remarks and future research are given in Sec. \ref{sec:Conclusions};
proofs and derivations are confined to the Appendices.

\emph{Notation} - Lower-case (resp. Upper-case) bold letters denote
vectors (resp. matrices), with $a_{n}$ (resp. $A_{n,m}$) representing
the $n$th (resp. the $(n,m)$th) element of the vector $\bm{a}$
(resp. matrix $\bm{A}$); upper-case calligraphic letters and braces
denote finite sets, with $[a:b]$, $a\leq b$, representing the set
$\{a,a+1,\ldots,b\}$; $\bm{I}_{N}$ denotes the $N\times N$ identity
matrix, while $\mathrm{diag}\left(\bm{t}\right)$ is a diagonal matrix
with diagonal equal to $\bm{t}$; $(\cdot)^{t}$,$\left\Vert \cdot\right\Vert _{2}$,
$\left\Vert \cdot\right\Vert _{F}$ and $\left\langle \cdot\right\rangle $
denote transpose, $\ell_{2}$ norm, Frobenius norm and inner product
operators, respectively; $\nabla_{\bm{t}}(\cdot)$ denotes the gradient
operator w.r.t. the vector $\bm{t}$; $\bm{e}^{j}(\bm{A},i)$ denotes
the unit eigenvector of a symmetric (and thus diagonalizable) matrix
$\bm{A}$ (of size $[r\times r]$) corresponding to the eigenvalue
$\lambda_{i}$, $i\in\mathcal{R}\triangleq\{1,\ldots r\}$, where
$\lambda_{s}>\lambda_{s+1}$, $s\in\mathcal{R}\backslash\{r\}$, and
$j\in\{-1,1\}$ denotes the sign ambiguity in the eigenvector formula;
$\arctan_{2}(\bm{x})$, $\bm{x}\in\mathbb{R}^{2}$, denotes the four-quadrant
inverse tangent with argument $\frac{x_{1}}{x_{2}}$; $P(\cdot)$
is used to denote probability mass functions (pmf) or probability
density functions (pdf), while $P(\cdot|\cdot)$ is the corresponding
conditional counterpart; $\mathcal{N}(\bm{\mu},\bm{\Sigma})$ denotes
a real normal distribution with mean vector $\bm{\mu}$ and covariance
matrix $\bm{\Sigma}$; finally the symbols $\rightarrow$, $\ni$,
$\sim$, and $\perp$ mean ``maps to'', ``such that'', \textquotedblleft{}distributed
as\textquotedblright{} and ``orthogonal'', respectively.

\section{System Model\label{sec:System-Model}}

\subsection{Motion Models description}

The system model is described graphically in Fig. \ref{fig:System Model}.
We assume that the target is observed by the platform at $n$ time
samples, i.e. $t\in\mathcal{T}\triangleq\{t_{1},\ldots,t_{n}\}$;
also we define the set of indices $\mathcal{I}\triangleq\{1,\ldots,n\}$.
In the following we will explicitly list all the assumptions made,
starting from the motion models of the platform and the target.

\textbf{Assumption I}: We assume that the target moves according to
a constant velocity (CV) motion model \cite{Bar-Shalom2002}. For
this reason we define $\bm{p}_{T}(t_{i})\triangleq\left[\begin{array}{cc}
\xi_{T}(t_{i}) & \eta_{T}(t_{i})\end{array}\right]^{t}$ and $\bm{v}_{T}\triangleq\left[\begin{array}{cc}
\dot{\xi}_{T} & \dot{\eta}_{T}\end{array}\right]^{t}$ as the position at $t_{i}$ and the (constant) velocity 2-D vector
of the target; $\xi$ and $\eta$ are used to denote the east and
north directions. Given the CV assumption, $\{\bm{p}_{T}(t_{1}),\bm{p}_{T}(t_{n})\}$
uniquely define the state of the target at $t_{i}\in\mathcal{T}$.
Therefore we stack them in $\bm{x}_{T}\triangleq\left[\begin{array}{cc}
\bm{p}_{T}(t_{1})^{t} & \bm{p}_{T}(t_{n})^{t}\end{array}\right]^{t}$, which represents the true target state vector, unknown at the platform
side. The target motion model has the explicit expression:
\begin{eqnarray}
\left[\begin{array}{c}
\xi_{T}(t_{i})\\
\eta_{T}(t_{i})
\end{array}\right] & = & \bm{p}_{T}(t_{1})+(t_{i}-t_{1})\cdot\bm{v}_{T}\label{eq:target motion model}\\
 & = & \bm{p}_{T}(t_{1})+\underbrace{\frac{(t_{i}-t_{1})}{(t_{n}-t_{1})}}_{\triangleq\alpha_{i}}\left[\bm{p}_{T}(t_{n})-\bm{p}_{T}(t_{1})\right],\quad t_{i}\in\mathcal{T}\label{eq:alternative target motion model}
\end{eqnarray}

\textbf{Assumption II}: We assume a platform moving according to a
``two-leg'' motion model; this requirement not only ensures observability
of the target from the platform point of view \cite{Nardone1981,Nardone1984},
but also represents the easiest trajectory that can be followed by
the platform. We denote $\bm{p}_{P}(t_{i})\triangleq\left[\begin{array}{cc}
\xi_{P}(t_{i}) & \eta_{P}(t_{i})\end{array}\right]^{t}$, $\bm{v}_{P,1}\triangleq\left[\begin{array}{cc}
\dot{\xi}_{P,1} & \dot{\eta}_{P,1}\end{array}\right]^{t}$ and $\bm{v}_{P,2}\triangleq\left[\begin{array}{cc}
\dot{\xi}_{P,2} & \dot{\eta}_{P,2}\end{array}\right]^{t}$, as the position at $t_{i}\in\mathcal{T}$, first-leg and second-leg
velocity vectors; also we group $\bm{x}_{P}\triangleq\left[\begin{array}{ccc}
\bm{p}_{P}(t_{1})^{t} & \bm{v}_{P,1}^{t} & \bm{v}_{P,2}^{t}\end{array}\right]^{t}$ into the platform state vector, representing the unknowns at the
target-friendly side. Note that $\bm{x}_{P}$ does not \emph{uniquely}
define the platform trajectory, since the turning time $t_{k}$ is
also needed.

\textbf{Assumption III}: Throughout this paper we will make the simplifying
assumption that $t_{k}$ is known at the target-friendly side. In
fact it is reasonable to assume that $t_{k}$ in practice will happen
nearly the middle of the observation interval, i.e., $t_{k}\approx\frac{t_{n}-t_{1}}{2}$,
in order to assure a good degree of observability%
\footnote{In fact, a turn at the beginning or the end of the observation interval
would result in a platform trajectory similar to a single leg (CV
model), thus leading to a nearly singular FIM.%
}. This assumption will be relaxed in Sec. \ref{sec:Simulation-Results},
where a sensitivity analysis with respect to (w.r.t.) the timing-uncertainty
on $t_{k}$ will be shown. Therefore, once $t_{k}$ is assumed to
be known, the platform motion model is explicitly described as 
\begin{equation}
\left[\begin{array}{c}
\xi_{P}(t_{i})\\
\eta_{P}(t_{i})
\end{array}\right]=\begin{cases}
\bm{p}_{P}(t_{1})+(t_{i}-t_{1})\bm{v}_{P,1} & t_{i}\in\mathcal{T},\,\ni\, t_{i}<t_{k}\\
\bm{p}_{P}(t_{1})+(t_{k}-t_{1})\bm{v}_{P,1}+(t_{i}-t_{k})\bm{v}_{P,2}\quad & t_{i}\in\mathcal{T},\,\ni\, t_{i}\geq t_{k}
\end{cases}\label{eq:Platform_Motion_Model}
\end{equation}
Note that here the magnitudes of $\bm{v}_{P,1}$ and $\bm{v}_{P,2}$
are arbitrary. As shown later, unique identifiability of the platform
trajectory (our goal) requires these magnitudes to be the same, i.e.,
the platform speed should be constant.

\begin{center}
\begin{figure}
\centering{}\includegraphics[clip,scale=0.6]{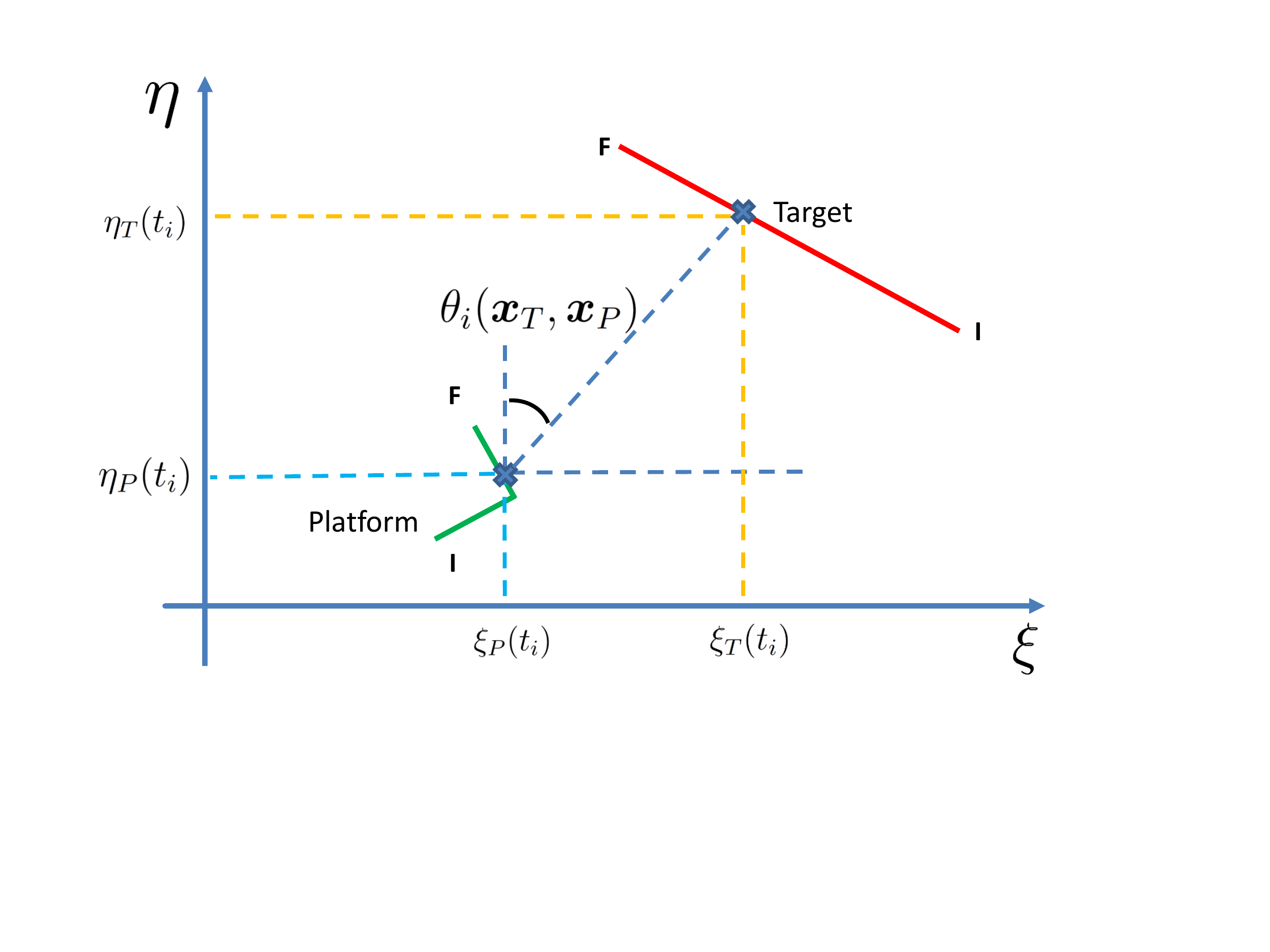}\caption{System model considered for our application. The crosses refer to
the bearing measurement at snapshot $t_{i}\in\mathcal{T}$.\label{fig:System Model}}
\end{figure}

\par\end{center}

\subsection{ML-PDA statistical assumptions and formulation}

The statistical assumptions on the measurements are summarized as
follows.

\textbf{Assumption I}: The bearing (true\textemdash{}originated from
the target) measurement $\kappa_{i}$, collected by the platform at
$t_{i}$, follows the model
\begin{eqnarray}
\kappa_{i} & = & \theta_{i}(\bm{x}_{T},\bm{x}_{P})+n_{i}=\arctan\left(\frac{\xi_{T}(t_{i})-\xi_{P}(t_{i})}{\eta_{T}(t_{i})-\eta_{P}(t_{i})}\right)+n_{i}\label{eq:observation_model_bearing-1}
\end{eqnarray}
where $\theta_{i}(\bm{x}_{T},\bm{x}_{P})$ denotes the noise-free
bearing (we stress the dependence on both platform and target state
vectors) and $n_{i}\sim\mathcal{N}(0,\sigma_{\theta}^{2})$. For notational
convenience we also define here the range $r_{i}(\bm{x}_{T},\bm{x}_{P})$
as
\begin{equation}
r_{i}(\bm{x}_{T},\bm{x}_{P})\triangleq\left\Vert \bm{p}_{T}(t_{i})-\bm{p}_{P}(t_{i})\right\Vert _{2}\label{eq:range_definition-1}
\end{equation}

\textbf{Assumption II}: We assume, as in realistic environments, that
a passive sonar at $t_{i}\in\mathcal{T}$ collects a set of measurements
$\bm{z}(i)$, due to clutter and non-perfect detection. More specifically,
we have 
\begin{equation}
\bm{z}(i)\triangleq\{z_{j}(i)\}_{j=1}^{m_{i}}\label{eq:set_measurements_MOU}
\end{equation}
where $m_{i}$ denotes the number of collected measurements at $t_{i}\in\mathcal{T}$.
The statistical assumptions over the set in Eq. (\ref{eq:set_measurements_MOU})
are: ($i$) the true measurement $\kappa_{i}$ can be detected at
most only once, with probability $P_{D}$; ($ii$) the number of false
measurements at $t_{i}\in\mathcal{T}$ follows a known probability
mass function $\mu_{F}(\cdot)$, given by a Poisson law with known
expected number of false alarms per unit of volume $\lambda$; therefore
the false measurements are distributed uniformly and independently
in the surveillance region (in the bearing space).

\textbf{Assumption III}: We assume conditional mutual independence
among the sets of measurements, that is $P(\bm{z}(i_{1}),\bm{z}(i_{2})|\bm{x}_{T},\bm{x}_{P})=P(\bm{z}(i_{1})|\bm{x}_{T},\bm{x}_{P})P(\bm{z}(i_{2})|\bm{x}_{T},\bm{x}_{P})$,
$\forall i_{1}\neq i_{2}$. 

Under these assumptions and denoting $\bm{\breve{x}}_{T}$ (resp.
$\bm{\breve{x}}_{P}$) as the true target (resp. platform) state vector,
the ML-PDA estimate $\bm{\hat{x}}_{T}$ is obtained as
\begin{align}
\bm{\hat{x}}_{T} & \triangleq\arg\max_{\bm{x}_{T}}\prod_{i=0}^{n}P(\bm{z}(i)|\bm{x}_{T},\bm{\breve{x}}_{P})
\end{align}
where the likelihood of $\bm{x}_{T}$  is given by:
\begin{align}
P(\bm{z}(i)|\bm{x}_{T},\bm{\breve{x}}_{P}) & =u^{-m_{i}}(1-P_{D})\mu_{F}(m_{i})\nonumber \\
 & +\frac{u^{1-m_{i}}P_{D}\mu_{F}(m_{i}-1)}{m_{i}}\sum_{j=1}^{m_{i}}\frac{1}{\sqrt{2\pi}\sigma_{\theta}}\times\exp\left(-\frac{1}{2}\left(\frac{z_{j}(i)-\theta_{i}(\bm{x}_{T},\bm{\breve{x}}_{P})}{\sigma_{\theta}}\right)^{2}\right)
\end{align}

Note that $\bm{\breve{x}}_{P}$ is assumed known at the platform side.
It was shown numerically in \cite{Jauffret1990} that the covariance
matrix of the ML-PDA estimator \emph{essentially} attains the CRLB
and therefore it can be regarded  as an efficient estimator. For this
reason, the covariance matrix is approximated by the inverse of the
FIM given by

\begin{equation}
\bm{J}(\bm{x}_{T},\bm{x}_{P},\alpha_{\theta})\triangleq\alpha_{\theta}\underbrace{\sum_{i=0}^{n}\bm{\nabla}_{\bm{x}_{T}}(\theta_{i}(\bm{x}_{T},\bm{x}_{P}))\bm{\nabla}_{\bm{x}_{T}}^{t}(\theta_{i}(\bm{x}_{T},\bm{x}_{P}))}_{\triangleq\bm{J}_{u}(\bm{x}_{T},\bm{x}_{P})}\label{eq:FIM_definition}
\end{equation}
 where $\alpha_{\theta}$ is defined as%
\footnote{Note that $\alpha_{\theta}$ should not be confused with $\alpha_{i}=\frac{t_{i}-t_{1}}{t_{n}-t_{1}}$
defined in Eq. (\ref{eq:alternative target motion model}). %
} 
\begin{equation}
\alpha_{\theta}\triangleq\frac{q_{2}}{\sigma_{\theta}^{2}}
\end{equation}
with $q_{2}$ representing the IRF \cite{Jauffret1990,Bar-Shalom2011}.
Note that 
\begin{equation}
q_{2}=q_{2}(\lambda v_{g},P_{D},g)
\end{equation}
where $v_{g}$ and $g$ denote the volume of the validation region
and the gating threshold, respectively. It can be shown, after some
manipulations, that $\bm{\nabla}_{\bm{x}_{T}}(\theta_{i}(\bm{x}_{T},\bm{x}_{P}))$
has the explicit expression
\begin{equation}
\bm{\nabla}_{\bm{x}_{T}}(\theta_{i}(\bm{x}_{T},\bm{x}_{P}))=\frac{1}{r_{i}(\bm{x}_{T},\bm{x}_{P})}\left[\begin{array}{c}
(1-\alpha_{i})\cos(\theta_{i}(\bm{x}_{T},\bm{x}_{P}))\\
-(1-\alpha_{i})\sin(\theta_{i}(\bm{x}_{T},\bm{x}_{P}))\\
\alpha_{i}\cos(\theta_{i}(\bm{x}_{T},\bm{x}_{P}))\\
-\alpha_{i}\sin(\theta_{i}(\bm{x}_{T},\bm{x}_{P}))
\end{array}\right]\label{eq:gradient_vector_for_FIM}
\end{equation}

The FIM at the platform side is necessarily evaluated as $\bm{J}^{obs}\triangleq\bm{J}(\bm{\breve{x}}_{P},\bm{\hat{x}}_{T},\breve{\alpha}_{\theta})$,
where $\breve{\alpha}_{\theta}$ denotes the true $\alpha_{\theta}$,
known at the platform side. Given the results of the estimation process
at the platform side, that is $\{\bm{\hat{x}}_{T},\bm{J}^{obs}\}$,
our task can be summarized as follows.

\emph{We wish to identify the platform state, represented by $\bm{x}_{P}$,
by observing only the estimation results of the ML-PDA, that is }$\{\bm{\hat{x}}_{T},\bm{J}^{obs}\}$\emph{.
It is worth remarking that the unknowns of this deterministic problem
are represented by $\{\bm{\breve{x}}_{P},\breve{\alpha}_{\theta}\}$.
In fact, even if $\breve{\alpha}_{\theta}$ does not contribute to
specifying the platform trajectory, it has to be identified to solve
this task.}

The first important remark is that the identification problem is a
function only of $\bm{\hat{x}}_{T}$ rather than the true trajectory
$\bm{\breve{x}}_{T}$. This has an important consequence: \emph{the
identification of the platform trajectory does not depend on the true
target trajectory $\bm{\breve{x}}_{T}$; however we will show that
due to the sensitivity w.r.t. the true platform parameters, a  larger
 FIM matrix will lead to an easier identification in terms of local
optimization routines.}

\section{Objective Function Determination \label{sec:Inverse-Localization-search}}

As a starting point of the identification problem, it would be natural
to solve the non-linear equation 
\begin{equation}
\bm{J}(\bm{\hat{x}}_{T},\bm{x}_{P},\alpha_{\theta})=\bm{J}^{obs}\label{eq:nonlinear_system_unobservable}
\end{equation}
for the variables $\bm{x}_{P}$ and $\alpha_{\theta}$. However, as
stated by the following proposition, we will show that this system
is \emph{unobservable}, since there exists an infinite number of solutions
satisfying Eq. (\ref{eq:nonlinear_system_unobservable}).
\begin{prop}
If $\{\bm{x}_{P}^{*},\alpha_{\theta}^{*}\}$ is a solution of (\ref{eq:nonlinear_system_unobservable}),
then each $\{\bm{x}_{P}^{'},\alpha_{\theta}^{'}\}$ generated by the
subspace 
\begin{eqnarray}
\{\bm{x}_{P}^{'},\alpha_{\theta}^{'}\} & = & \{\beta\bm{x}_{P}^{*}+(1-\beta)\bm{\bar{x}}_{E},\beta^{2}\alpha_{\theta}^{*}\},\quad\beta\in\mathbb{R}\label{eq:Subspace of unobservability}\\
\bm{\bar{x}}_{E} & \triangleq & \left[\begin{array}{ccc}
\bm{\hat{p}}_{T}(t_{1})^{t} & \bm{\hat{v}}_{T}^{t} & \bm{\hat{v}}_{T}^{t}\end{array}\right]^{t}
\end{eqnarray}
 is also a solution of (\ref{eq:nonlinear_system_unobservable}).\label{prop:unobservability_statement}\end{prop}
\begin{IEEEproof}
The proof is given in Appendix \ref{sec:Appendix-unobservability-statement}.
\end{IEEEproof}
The above proposition states that \emph{no unambiguous identification
of the platform trajectory }$\bm{\breve{x}}_{P}$\emph{ is possible}
when the platform is following a trajectory according to Eq. (\ref{eq:Platform_Motion_Model})
with different speed in each leg. The explanation of this is given
by the fact that an affine combination (through $\beta$) of the platform
and the (augmented) target state vector would produce the same FIM
with an $\alpha_{\theta}$ scaled by $\beta^{2}$. At this point it
is worth stressing the \emph{difference} between this requirement
and the target estimation, \emph{in which the platform maneuver is
the only prerequisite }to ensure observability of the target by the
platform. 

However, when both legs of the platform trajectory are constrained
to have the same speed $s$, that is 
\begin{equation}
s\triangleq\left\Vert \bm{v}_{P,1}\right\Vert _{2}=\left\Vert \bm{v}_{P,2}\right\Vert _{2}\label{eq:equal_speed_two_legs}
\end{equation}
 we can show that the subspace described by Eq. (\ref{eq:Subspace of unobservability})
violates constraint (\ref{eq:equal_speed_two_legs}), if the specific
condition 
\begin{equation}
(\bm{v}_{P,1}-\bm{v}_{P,2})^{t}\bm{\hat{v}}_{T}\neq0\label{eq:orthogonality_condition_subspace}
\end{equation}
holds. The latter represents the condition under which the difference
vector of the two legs velocities is orthogonal to the velocity vector
of the target; such a trajectory would \emph{make the platform unobservable},
even though the constraint in Eq. (\ref{eq:equal_speed_two_legs})
is satisfied, and thus it should be intended as a \emph{stealthy trajectory}
achievable by the platform%
\footnote{Obviously, design of such platform trajectory would require in advance
the knowledge of the estimated target trajectory and thus it is not
performable in practice.%
}. To demonstrate this, let us consider the squared speed of the two
legs for a platform trajectory belonging to the subspace in Eq. (\ref{eq:Subspace of unobservability}):
\begin{eqnarray}
\left\Vert \bm{v}_{P,1}^{'}\right\Vert _{2}^{2} & = & \beta^{2}\left\Vert \bm{v}_{P,1}\right\Vert _{2}^{2}+(1-\beta)^{2}\left\Vert \hat{\bm{v}}_{T}\right\Vert _{2}^{2}+2\beta(1-\beta)\hat{\bm{v}}_{T}^{t}\bm{v}_{P,1}\\
\left\Vert \bm{v}_{P,2}^{'}\right\Vert _{2}^{2} & = & \beta^{2}\left\Vert \bm{v}_{P,2}\right\Vert _{2}^{2}+(1-\beta)^{2}\left\Vert \hat{\bm{v}}_{T}\right\Vert _{2}^{2}+2\beta(1-\beta)\hat{\bm{v}}_{T}^{t}\bm{v}_{P,2}
\end{eqnarray}
If we constrain the true platform trajectory to keep constant speed
during the two legs, evaluating the difference $\left\Vert \bm{v}_{P,1}^{'}\right\Vert _{2}^{2}-\left\Vert \bm{v}_{P,2}^{'}\right\Vert _{2}^{2}$
leads to 
\begin{equation}
\left\Vert \bm{v}_{P,1}^{'}\right\Vert _{2}^{2}-\left\Vert \bm{v}_{P,2}^{'}\right\Vert _{2}^{2}=2\beta(1-\beta)\hat{\bm{v}}_{T}^{t}\left(\bm{v}_{P,1}-\bm{v}_{P,2}\right)\label{eq:subspace_violation}
\end{equation}
 From inspection of Eq. (\ref{eq:subspace_violation}), it is apparent
that a platform described by Eq. (\ref{eq:Subspace of unobservability})
keeps a constant speed in the two legs only if: ($i$) $\beta=1$,
i.e. the trajectory considered coincides with $\bm{x}_{P}^{*}$; ($ii$)
$\beta=0$, which represents a degenerate platform trajectory and
thus it can be excluded; ($iii$) $\hat{\bm{v}}_{T}\perp\left(\bm{v}_{P,1}-\bm{v}_{P,2}\right)$.
For this reason, under the assumptions in Eqs. (\ref{eq:equal_speed_two_legs})
and (\ref{eq:orthogonality_condition_subspace}), identification of
the platform trajectory is possible. More specifically, the constraint
in Eq. (\ref{eq:equal_speed_two_legs}) represents a \emph{necessary
condition} for \emph{observability} of the platform trajectory. On
the basis of this constraint we \emph{define} a new platform-state
vector $\bm{x}_{P}^{s}$ (and we denote the true platform-state vector
as $\bm{\breve{x}}_{P}^{s}$) as follows 
\begin{eqnarray}
\bm{x}_{P}^{s} & \triangleq & \left[\begin{array}{ccccc}
\xi_{P}(t_{1}) & \eta_{P}(t_{1}) & s & \phi_{1} & \phi_{2}\end{array}\right]^{t}\\
\phi_{i} & \triangleq & \arctan_{2}\left(\bm{v}_{P,i}\right),\qquad i\in\{1,2\}
\end{eqnarray}
Thus Eq. (\ref{eq:nonlinear_system_unobservable}) becomes 
\begin{equation}
\bm{J}(\bm{x}_{P}^{s},\alpha_{\theta})=\bm{J}^{obs}\label{eq:nonlinear_system}
\end{equation}
with unknowns $\bm{x}_{P}^{s}$ and $\alpha_{\theta}$ (starting from
here we drop the dependence on $\bm{\hat{x}}_{T}$ to keep the notation
simple). In general, the non-linear system described by Eq. (\ref{eq:nonlinear_system})
can still admit multiple solutions, since there is no theoretical
proof that the set of constraints in Eqs. (\ref{eq:equal_speed_two_legs})
and (\ref{eq:orthogonality_condition_subspace}) is also a sufficient
condition for observability of $\{\bm{x}_{P}^{s},\alpha_{\theta}\}$.
This is because proving that $\{\bm{x}_{P}^{s},\alpha_{\theta}\}\rightarrow\bm{J}(\bm{x}_{P}^{s},\alpha_{\theta})$
is a \emph{one-to-one mapping} is an extremely difficult task. Nonetheless,
we will show, through simulations in Sec. \ref{sec:Simulation-Results},
that this property seems to be satisfied and that $\bm{x}_{P}^{s}$
can be identified.

To solve Eq. (\ref{eq:nonlinear_system}) in an efficient way we consider
the search for the minimum of the square of the Frobenius norm $\mathcal{F}(\bm{x}_{P}^{s},\alpha_{\theta})$,
namely,
\begin{eqnarray}
\{\bm{\hat{x}}_{P}^{s},\hat{\alpha}_{\theta}\} & = & \arg\min_{\{\bm{x}_{P}^{s},\alpha_{\theta}\}}\underbrace{\left\Vert \bm{J}^{obs}-\bm{J}(\bm{x}_{P}^{s},\alpha_{\theta})\right\Vert _{F}^{2}}_{\triangleq\mathcal{F}(\bm{x}_{P}^{s},\alpha_{\theta})}\label{eq:Frobn_minimization}
\end{eqnarray}

It is easy to see that the global minimum (corresponding to a zero
value) of Eq. (\ref{eq:Frobn_minimization}) corresponds to the solution
of Eq. (\ref{eq:nonlinear_system}). Although the criterion of Eq.
(\ref{eq:Frobn_minimization}) appears as arbitrary (in fact other
matrix distance norms can be considered), there is an important reason
behind this choice: we will show in the following that $\mathcal{F}(\bm{x}_{P}^{s},\alpha_{\theta})$
(see (\ref{eq:weighted_nonlinear_LS_separated})) can be expressed
in terms of a \emph{weighted-square-distance}, and an important property
of weighted non-linear least squares problems can be exploited \cite{Kay1993}.

The first step to express Eq. (\ref{eq:Frobn_minimization}) in terms
of a convenient weighted-least squares problem is to search for independent
entries of $\bm{J}(\bm{x}_{P}^{s},\alpha_{\theta})$. The following
Lemma will be used. 
\begin{lem}
The FIM $\bm{J}(\bm{x}_{P}^{s},\alpha_{\theta})$ has only $9$ independent
entries.\label{lem:Fisher_Matrix_9elements}\end{lem}
\begin{IEEEproof}
The proof is given in Appendix \ref{sec:Proof-of-9 elements FIM}.
\end{IEEEproof}
Exploiting Lemma \ref{lem:Fisher_Matrix_9elements}, we can express
Eq. (\ref{eq:Frobn_minimization}) in terms of a weighted-square distance,
as stated by the following proposition.
\begin{prop}
The square of the norm in Eq. (\ref{eq:Frobn_minimization}) can be
equivalently written in the form\label{prop:weighted_squared_matrix_FIM}
\begin{equation}
\mathcal{F}(\bm{x}_{P}^{s},\alpha_{\theta})=\left[\bm{j}^{obs}-\bm{j}(\bm{x}_{P}^{s},\alpha_{\theta})\right]^{t}\bm{W}\left[\bm{j}^{obs}-\bm{j}(\bm{x}_{P}^{s},\alpha_{\theta})\right]\label{eq:weighted_nonlinear_LS}
\end{equation}
where $\bm{j}^{obs}\in\mathbb{R}^{9}$ and $\bm{j}(\bm{x}_{P}^{s},\alpha_{\theta})\in\mathbb{R}^{9}$
are obtained by stacking the independent components of $\bm{J}(\bm{x}_{P}^{s},\alpha_{\theta})$
and $\bm{J}^{obs}$, respectively and $\bm{W}$ is a diagonal weighting
matrix, defined as follows:
\begin{equation}
\bm{W}\triangleq\mathrm{diag}\left(\left[\begin{array}{ccccccccc}
1 & 1 & 1 & 1 & 2 & 2 & 4 & 2 & 2\end{array}\right]^{t}\right)
\end{equation}
\end{prop}
\begin{IEEEproof}
The proof is given in Appendix \ref{sec:Appendix-weighted squared distance}. 
\end{IEEEproof}
Note that $\bm{j}(\bm{x}_{P}^{s},\alpha_{\theta})$ retains the same
factorization as $\bm{J}(\bm{x}_{P}^{s},\alpha_{\theta})$, that is,
$\bm{j}(\bm{x}_{P}^{s},\alpha_{\theta})=\alpha_{\theta}\bm{j}_{u}(\bm{x}_{P}^{s})$,
where $\bm{j}_{u}(\bm{x}_{P}^{s})$ is defined accordingly to Eq.
(\ref{eq:FIM_definition}). Hence Eq. (\ref{eq:weighted_nonlinear_LS})
can be rewritten as
\begin{equation}
\mathcal{F}(\bm{x}_{P}^{s},\alpha_{\theta})=\left[\bm{j}^{obs}-\alpha_{\theta}\bm{j}_{u}(\bm{x}_{P}^{s})\right]^{t}\bm{W}\left[\bm{j}^{obs}-\alpha_{\theta}\bm{j}_{u}(\bm{x}_{P}^{s})\right]\label{eq:weighted_nonlinear_LS_separated}
\end{equation}

Note that the minimization of the objective function in Eq. (\ref{eq:weighted_nonlinear_LS_separated})
is in the standard form of non-linear weighted least squares \cite{Kay1993}.
Additionally in this case the non-linear weighted square distance
is linear in some of the parameters to be estimated, in this case
$\alpha_{\theta}$, and thus this non-linear problem can be solved
in a reduced dimension space; the details are given by the following
proposition.
\begin{prop}
The minimization in the 6-dimensional space of the objective $\mathcal{F}(\bm{x}_{P}^{s},\alpha_{\theta})$
is equivalent to maximization of the objective $\mathcal{G}(\bm{x}_{P}^{s})$,
defined as \label{prop:new_objective_function}
\begin{eqnarray}
\mathcal{G}(\bm{x}_{P}^{s}) & \triangleq & \left\langle \bm{j}^{obs},\bm{c}(\bm{x}_{P}^{s})\right\rangle ^{2}\label{eq:final_obj_function}\\
\bm{c}(\bm{x}_{P}^{s}) & \triangleq & \frac{\bm{W}\bm{j}_{u}(\bm{x}_{P}^{s})}{\left[\bm{j}_{u}(\bm{x}_{P}^{s})^{t}\bm{W}\bm{j}_{u}(\bm{x}_{P}^{s})\right]^{\nicefrac{1}{2}}}
\end{eqnarray}
\end{prop}
\begin{IEEEproof}
The proof is given in Appendix \ref{sec:Appendix_new_objective_function_demonstration}.
\end{IEEEproof}
The maximization of $\mathcal{G}(\bm{x}_{P}^{s})$ has the advantage
of reducing the search space from $\mathbb{R}^{6}$ to $\mathbb{R}^{5}$.
The exclusion of $\alpha_{\theta}$ also allows a search \emph{only}
in the subspace of variables determining the platform trajectory.
At this point some observations on the objective function in Eq. (\ref{eq:final_objective_function_1})
are in order:
\begin{itemize}
\item The function $\mathcal{G}(\bm{x}_{P}^{s})$ is highly non-linear in
$\bm{x}_{P}^{s}$, therefore no closed form solution to optimization
of  Eq. (\ref{eq:final_obj_function}) exists; thus numerical optimization
procedures need to be used;
\item Local optimization routines can get stuck in local maxima or other
(non-equilibrium) points%
\footnote{In fact, several local optimization routines are not based on the
evaluation of the gradient vector of the objective function at each
iteration, thus a ``null gradient'' condition is not ensured when
a stopping condition is met.%
}.
\end{itemize}
An important issue in local optimization routines that determines
their success, is the choice of a good initial guess. In the following
we will suggest a good start, based on the FIM and platform-target
geometry.

\section{``Geometry-driven'' FIM-aided initial estimate choice\label{sec:Initial Guess Choice}}

In this section we show how a good initial estimate, denoted as $\hat{\bm{x}}{}_{P}^{s,0}$,
can be chosen to help the convergence of the local optimization routines.
To accomplish this task define
\begin{equation}
\bm{p}_{P}\triangleq\begin{cases}
\left[\begin{array}{ccc}
\bm{p}_{P}(t_{1})^{t} & \bm{p}_{P}(t_{k})^{t} & \bm{p}_{P}(t_{n})^{t}\end{array}\right]^{t}\\
\ni\quad\frac{\left\Vert \bm{p}_{P}(t_{k})-\bm{p}_{P}(t_{1})\right\Vert _{2}}{t_{k}-t_{1}}=\frac{\left\Vert \bm{p}_{P}(t_{n})-\bm{p}_{P}(t_{k})\right\Vert _{2}}{t_{n}-t_{k}}
\end{cases}\label{eq:p_p- local optimization procedure}
\end{equation}
Since it is easy to show that there is a \emph{one-to-one mapping}
between $\bm{p}_{P}$ and $\bm{x}{}_{P}^{s}$, we will search for
a $\bm{\hat{p}}_{P}^{0}$ close to $\bm{\breve{p}}_{P}$ (even though
the latter is not known), as the equivalent input to the local optimization
routine. For this reason, we first will find good approximations $\bm{\hat{p}}_{P}^{0}(t_{1})$
and $\bm{\hat{p}}_{P}^{0}(t_{n})$. After this, we will give the details
on how to find $\bm{\hat{p}}_{P}^{0}(t_{k})$, under the constraint
of Eq. (\ref{eq:p_p- local optimization procedure}). The following
considerations are based on the assumption that $\breve{\alpha}_{\theta}$
is known; at the end of the section we will remove this restriction.
In the following, for the sake of simplicity, we will use the short-hand
notations $r_{i}\triangleq r_{i}(\bm{\hat{x}}_{T},\bm{\breve{x}}_{P}^{s})$
for the range, $\theta_{i}\triangleq\theta_{i}(\bm{\hat{x}}_{T},\bm{\breve{x}}_{P}^{s})$
for the bearing, $\bm{i}_{i}\triangleq\left[\begin{array}{cc}
\sin\left(\theta_{i}\right) & \cos\left(\theta_{i}\right)\end{array}\right]^{t}$ for the bearing unit vector, and we will drop the zero superscript
in $\bm{\hat{p}}_{P}^{0}$ and $\bm{\hat{p}}_{P}^{0}(t_{i})$.

\subsection{Choice of \textmd{\normalsize{$\hat{\bm{p}}_{P}(t_{1})$ and $\hat{\bm{p}}_{P}(t_{n})$}} }

It can be easily shown that $\{\bm{\breve{p}}_{P}(t_{1}),\bm{\breve{p}}_{P}(t_{n})\}$
can be expressed in the form:
\begin{eqnarray}
\bm{\breve{p}}_{P}(t_{1}) & = & \bm{\hat{p}}_{T}(t_{1})+r_{1}\bm{i}_{1}\label{eq:ppt0_pptn_vector_range_form}\\
\bm{\breve{p}}_{P}(t_{n}) & = & \bm{\hat{p}}_{T}(t_{n})+r_{n}\bm{i}_{n}
\end{eqnarray}

In order to obtain good estimates $\{\bm{\hat{p}}_{P}(t_{1}),\bm{\hat{p}}_{P}(t_{n})\}$
we need to find good approximations $\{\hat{r}_{1},\hat{r}_{n}\}$
and $\{\hat{\bm{i}}_{1},\bm{\hat{i}}_{n}\}$. These issues can be
tackled separately.

\subsubsection*{Choice of $\{\hat{r}_{1},\hat{r}_{n}\}$}

A coarse approximation of the ranges $r_{1}$ and $r_{n}$ is given
by:
\begin{equation}
\hat{r}_{1}\triangleq\sqrt{\frac{\breve{\alpha}_{\theta}\sum_{i=1}^{n}(1-\alpha_{i})^{2}}{\mathrm{tr}\left(\bm{J}^{obs}[1,1]\right)}};\quad\hat{r}_{n}\triangleq\sqrt{\frac{\breve{\alpha}_{\theta}\sum_{i=1}^{n}\alpha_{i}{}^{2}}{\mathrm{tr}\left(\bm{J}^{obs}[2,2]\right)}}\label{eq:sec_approx_ri}
\end{equation}
where $\bm{J}^{obs}[m,n]$ represents the $(m,n)$th $[2\times2]$
block matrix of $\bm{J}^{obs}$, whose explicit expression in block
form (exploiting the derivation in Appendix \ref{sec:Proof-of-9 elements FIM})
is given as follows:
\begin{align}
\bm{J}^{obs} & =\left[\begin{array}{cc}
\bm{J}^{obs}[1,1] & \bm{J}^{obs}[1,2]\\
\bm{J}^{obs}[2,1] & \bm{J}^{obs}[2,2]
\end{array}\right]\\
 & =\alpha_{\theta}\left[\begin{array}{cc}
\sum_{i=1}^{n}(1-\alpha_{i})^{2}\bm{\breve{y}}_{i}\bm{\breve{y}}_{i}^{t} & \sum_{i=1}^{n}\alpha_{i}(1-\alpha_{i})\bm{\breve{y}}_{i}\bm{\breve{y}}_{i}^{t}\\
\sum_{i=1}^{n}\alpha_{i}(1-\alpha_{i})\bm{\breve{y}}_{i}\bm{\breve{y}}_{i}^{t} & \sum_{i=1}^{n}\alpha_{i}{}^{2}\bm{\breve{y}}_{i}\bm{\breve{y}}_{i}^{t}
\end{array}\right]
\end{align}
where 
\begin{equation}
\bm{\breve{y}}_{i}\triangleq\frac{1}{r_{i}(\bm{\hat{x}}_{T},\bm{\breve{x}}_{P}^{s})}\left[\begin{array}{cc}
\cos(\theta_{i}(\bm{\hat{x}}_{T},\bm{\breve{x}}_{P}^{s})) & -\sin(\theta_{i}(\bm{\hat{x}}_{T},\bm{\breve{x}}_{P}^{s}))\end{array}\right]^{t}
\end{equation}
The term $\mathrm{tr}\left(\bm{J}^{obs}[1,1]\right)$ (resp. $\mathrm{tr}\left(\bm{J}^{obs}[2,2]\right)$)
represents the sum of the eigenvalues of the FIM of $\bm{p}_{T}(t_{1})$
(resp. $\bm{p}_{T}(t_{n})$), that is, the sum of the square of the
semiaxes of the corresponding ellipse. The terms in the numerators
of Eq. (\ref{eq:sec_approx_ri}) are correction factors which avoid
biased estimates $\hat{r}_{1}$ and $\hat{r}_{n}$\emph{.} The derivation
of Eq. (\ref{eq:sec_approx_ri}) is given in Appendix \ref{sec: Appendix_ choice r_hat i}.

It is worth remarking that Eq. (\ref{eq:sec_approx_ri}) \emph{represents
a rough approximation} of $\{r_{1},r_{n}\}$; however, as we will
show in Sec. \ref{sec:Simulation-Results}, the accuracy obtained
is sufficient to determine values close to $\bm{\breve{p}}_{P}(t_{1})$
and $\bm{\breve{p}}_{P}(t_{n})$ in all the practical scenarios considered.

\subsubsection*{Choice of $\{\bm{\hat{i}}_{1},\bm{\hat{i}}_{n}\}$}

By defining $\bm{C}^{obs}\triangleq\left(\bm{J}^{obs}\right)^{-1}$
and denoting $\bm{C}^{obs}[\ell,m]$ as the $(\ell,m)$th $[2\times2]$
block matrix of $\bm{C}^{obs}$ we have that:

\begin{equation}
\bm{\hat{i}}_{1}^{p}\triangleq\bm{e}^{p}\left(\bm{C}^{obs}[1,1];1\right),\qquad\qquad\bm{\hat{i}}_{n}^{q}\triangleq\bm{e}^{q}\left(\bm{C}^{obs}[2,2];1\right);\qquad p,q\in\{-1,1\}\label{eq:approx_i_i}
\end{equation}
Note that $\bm{C}^{obs}[1,1]$ (resp. $\bm{C}^{obs}[2,2]$) is a lower
bound of the covariance matrix of any unbiased estimator of $\bm{p}_{T}(t_{1})$
(resp. $\bm{p}_{T}(t_{n})$) and thus the eigenvector corresponding
to the largest eigenvalue represents the axis of maximum uncertainty,
i.e., the one along the range between the platform and the target.
The indices $\{p,q\}$ underline the incomplete information about
$\{\bm{i}_{1},\bm{i}_{n}\}$ contained in the FIM, which intrinsically
leads to a sign ambiguity (because if $\bm{e}$ is an eigenvector,
so is $-\bm{e}$). The derivation of Eq. (\ref{eq:approx_i_i}) is
given in Appendix \ref{sec:Appendix_choice-of_i0-in}.

\begin{figure}
\centering{}\includegraphics[width=0.4\paperwidth]{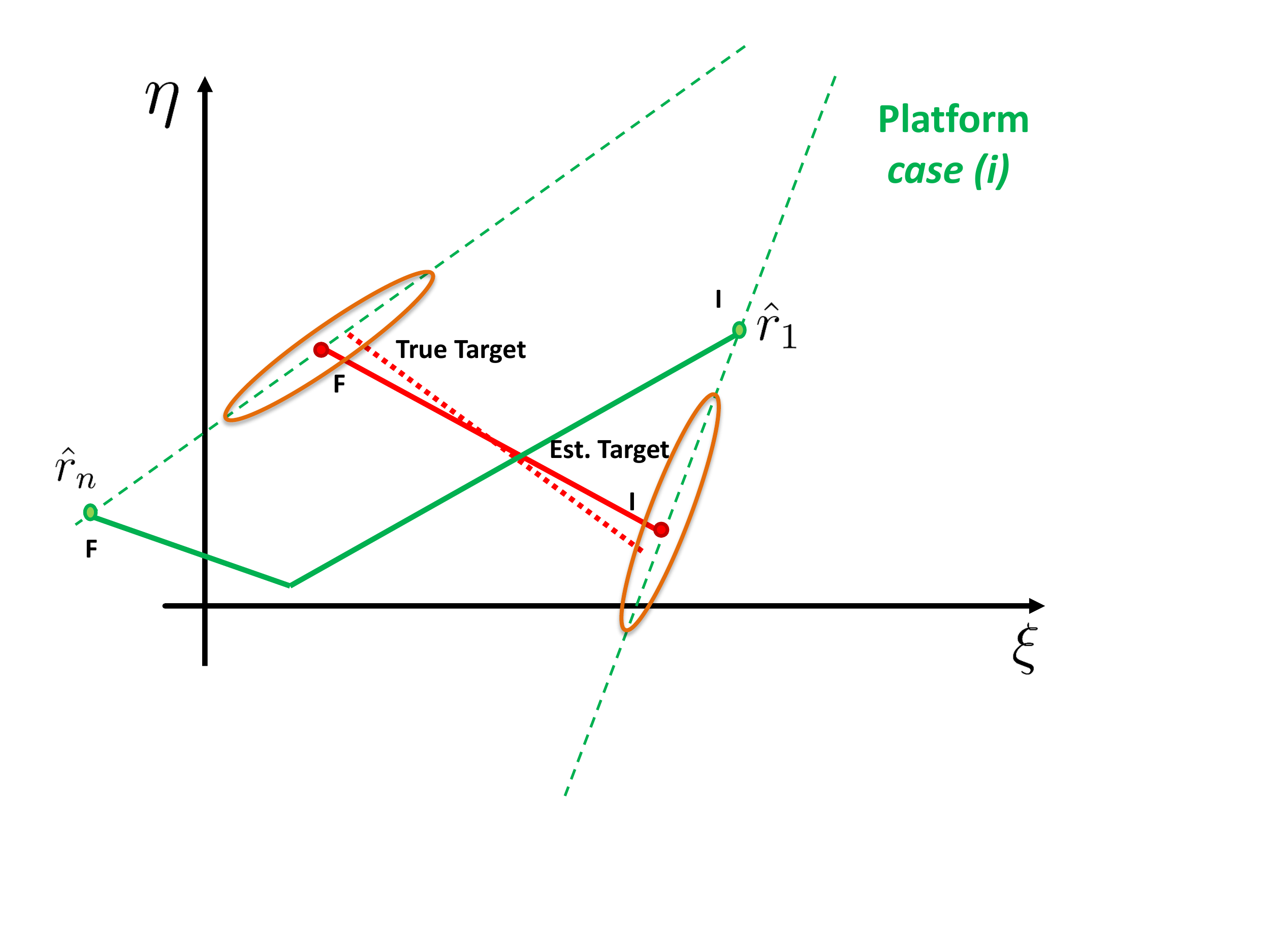}\includegraphics[width=0.4\paperwidth]{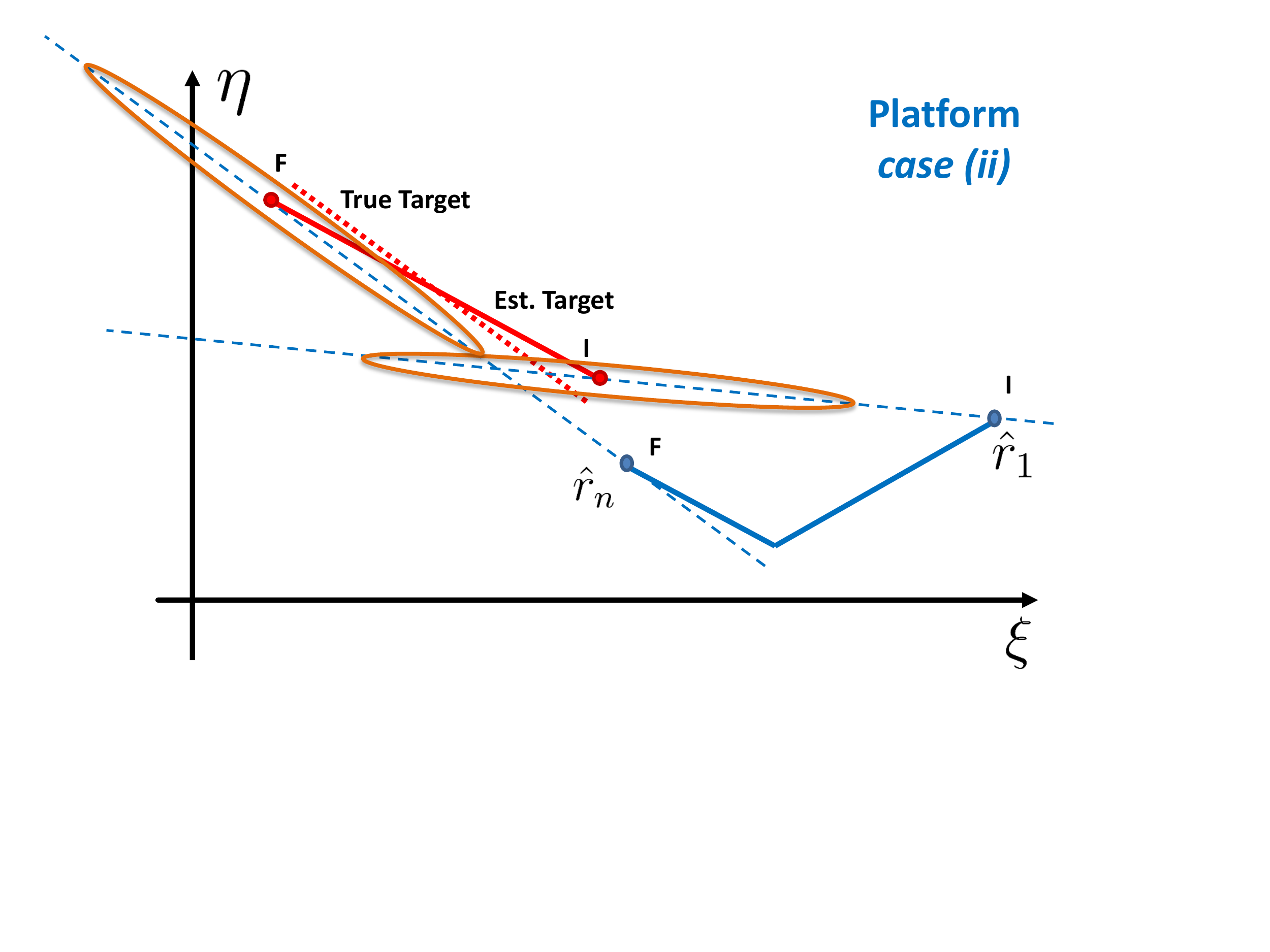}\caption{Graphical description of Eq. (\ref{eq:only_twoeigenvector_couples_assumpt}):
cases ($i$) and ($ii$) satisfying it. \label{fig:Intersection-Eigenvectors}}
\end{figure}

It is apparent that the sign ambiguity in Eq. (\ref{eq:approx_i_i})
would lead to four possible pairs $\{\bm{\hat{i}}_{1}^{p},\bm{\hat{i}}_{n}^{q}\}$.
However, in practical scenarios the sign-ambiguity leads to only two
pairs if we assume that the following vector equation \emph{has no
solution}:
\begin{equation}
\zeta\cdot\bm{\hat{p}}_{T}(t_{1})+(1-\zeta)\cdot\bm{\hat{p}}_{T}(t_{n})=\bm{p}_{P}(t_{i});\quad\forall t_{i}\in\mathcal{T},\,\zeta\in\mathbb{R}\label{eq:only_twoeigenvector_couples_assumpt}
\end{equation}
Eq. (\ref{eq:only_twoeigenvector_couples_assumpt}) admits solutions
only in the following cases%
\footnote{Although Eq. (\ref{eq:only_twoeigenvector_couples_assumpt}) refers
to $\bm{\hat{p}}_{T}(t_{i})$, the same applies to $\bm{p}_{T}(t_{i})$
under the assumption that the platform obtains a reasonably ``confident''
estimate of the target.%
}: ($i$) the platform trajectory and the target trajectory have at
least one crossing point; ($ii$) the platform is crossing the line
representing the direction of the CV target trajectory, determined
by its velocity vector $\bm{\hat{v}}_{T}$. These scenarios are graphically
depicted in Fig. \ref{fig:Intersection-Eigenvectors}. While the former
case represents a totally unrealistic scenario (the platform and the
target would be too near), it can be shown that the latter is of little
interest, since such a platform trajectory would lead to poor observability
of the target \cite{Nardone1984}. 

Therefore the only two admissible pairs are obtained as follows: let
us define a vector $\{\bm{u}\in\mathbb{R}^{2}:\bm{u}\perp\bm{\hat{v}}_{T},\left\Vert \bm{u}\right\Vert _{2}=1\}$
and take the pairs $\{\hat{\bm{i}}_{1}^{p},\hat{\bm{i}}_{n}^{q}\}$
satisfying
\begin{equation}
\mathrm{sign}\left\langle \hat{\bm{i}}_{1}^{p},\bm{u}\right\rangle =\mathrm{sign}\left\langle \hat{\bm{i}}_{n}^{q},\bm{u}\right\rangle ,\quad p,q\in\{-1,1\}\label{eq:i_i_feasibility}
\end{equation}
 In the following we denote as $\{\bm{\hat{i}}_{1}^{g},\bm{\hat{i}}_{n}^{g}\}$,
$g\in\{1,2\}$, the resulting two pairs. Thus, exploiting Eqs. (\ref{eq:sec_approx_ri}),
(\ref{eq:approx_i_i}) and (\ref{eq:i_i_feasibility}) we obtain:
\begin{eqnarray}
\hat{\bm{p}}_{P}^{g}(t_{1}) & = & \bm{\hat{p}}_{T}(t_{1})+\hat{r}_{1}\cdot\bm{\hat{i}}_{1}^{g},\qquad g\in\{1,2\}\label{eq:initial_guess_ppt0}\\
\hat{\bm{p}}_{P}^{g}(t_{n}) & = & \bm{\hat{p}}_{T}(t_{n})+\hat{r}_{n}\cdot\bm{\hat{i}}_{n}^{g},\qquad g\in\{1,2\}\label{eq:initial_guess_pptn}
\end{eqnarray}

\subsection{Choice of turning position vector \textmd{\normalsize{$\hat{\bm{p}}_{P}(t_{k})$}}}

As pointed out previously, Eqs. (\ref{eq:initial_guess_ppt0}) and
(\ref{eq:initial_guess_pptn}) define two pairs $\{\bm{\hat{p}}_{P}^{g}(t_{1}),\bm{\hat{p}}_{P}^{g}(t_{n})\}$,
$g\in\{1,2\}$. Hence for each $\{\bm{\hat{p}}_{P}^{g}(t_{1}),\bm{\hat{p}}_{P}^{g}(t_{n})\}$
a corresponding $\bm{\hat{p}}_{P}^{g}(t_{k})$ needs to be computed.
In this case it can be shown that $\bm{\hat{p}}_{P}^{g}(t_{k})$ cannot
be chosen only relying on the geometric properties of the FIM, as
opposed to $\bm{\hat{p}}_{P}(t_{1})$ and $\bm{\hat{p}}_{P}(t_{n})$.
Rather, a $\pm\frac{\pi}{2}$ platform turn initial assumption is
made and the turn-sign ambiguity is solved exploiting the coarse information
in the FIM. By defining $t_{m}\triangleq\arg\min_{t_{i}\in\mathcal{T}}\left\Vert t_{i}-\frac{t_{n}-t_{1}}{2}\right\Vert _{2}$,
such a vector is obtained as (we drop the superscript $g$)
\begin{eqnarray}
\bm{\hat{p}}_{P}(t_{k}) & = & \bm{\rho}_{q}(t_{k})\label{eq:ambiguity_solving_pptk}\\
\bm{\rho}_{\ell}(t_{k}) & \triangleq & \hat{\bm{p}}_{P}(t_{1})+\frac{(t_{k}-t_{1})}{(t_{n}-t_{k})}\left\Vert \hat{\bm{p}}_{P}(t_{n})-\hat{\bm{p}}_{P}(t_{1})\right\Vert _{2}\cos\left(\nu\right)\left[\begin{array}{c}
\sin\left(\psi_{\ell}\right)\\
\cos\left(\psi_{\ell}\right)
\end{array}\right]\\
\psi_{\ell} & \triangleq & \arctan_{2}\left(\hat{\bm{p}}_{P}(t_{n})-\hat{\bm{p}}_{P}(t_{1})\right)+\ell\cdot\left(\frac{\pi}{2}-\nu\right)\\
\nu & \triangleq & \arctan\left(\frac{t_{k}-t_{1}}{t_{n}-t_{k}}\right)\\
q & \triangleq & \arg\min_{\ell\in\{-1,1\}}\left\Vert \tilde{\bm{p}}{}_{P}(t_{m})-\bm{\rho}_{\ell}(t_{m})\right\Vert _{2}
\end{eqnarray}
where $\bm{\rho}_{\ell}(t_{m})$ is the position vector at $t_{m}$
of the two-leg trajectory described by $\{\hat{\bm{p}}_{P}(t_{1}),\bm{\rho}_{\ell}(t_{k}),\hat{\bm{p}}_{P}(t_{n})\}$,
and $\tilde{\bm{p}}{}_{P}(t_{m})$ is given by
\begin{align}
\tilde{\bm{p}}{}_{P}(t_{m}) & \triangleq\bm{\hat{p}}_{T}(t_{m})+\tilde{r}{}_{m}\cdot\bm{\tilde{i}}_{m}\\
\tilde{r}_{m} & \triangleq\sqrt{\frac{\breve{\alpha}_{\theta}\sum_{i=1}^{n}\alpha_{i}(1-\alpha_{i})}{\mathrm{tr}\left(\bm{J}^{obs}[1,2]\right)}}\\
\tilde{\bm{i}}_{m} & \triangleq\bm{e}^{q}\left(\bm{J}^{obs}[1,2];1\right)\\
q & \ni\mathrm{sign}\left\langle \bm{e}^{j}\left(\bm{J}^{obs}[1,2];1\right),\bm{u}\right\rangle =\mathrm{sign}\left\langle \hat{\bm{i}}_{1},\bm{u}\right\rangle ,\quad j\in\{-1,1\}.\label{eq:coarse_guess_pptk}
\end{align}

The complete derivation of the selection of $\bm{\hat{p}}_{p}^{g}(t_{k})$
is given in Appendix \ref{sec:Appendix_choice of pptk}.

\subsection{Remarks on $\alpha_{\theta}$\label{sub:Remarks_alpha_theta}}

The presented method computes $\hat{\bm{p}}_{P}^{g}=\left[\begin{array}{ccc}
\hat{\bm{p}}_{P}^{g}(t_{1})^{t} & \hat{\bm{p}}_{P}^{g}(t_{k})^{t} & \hat{\bm{p}}_{P}^{g}(t_{n})^{t}\end{array}\right]^{t}$, $g\in\{1,2\}$, under the assumption that $\breve{\alpha}_{\theta}$
is known at the target-friendly side. However, since $\breve{\alpha}_{\theta}$
is known exactly only at the platform side, we will replace $\breve{\alpha}_{\theta}$
with the variable $\alpha_{\theta}$ in Eq. (\ref{eq:sec_approx_ri}),
thus leading to $\hat{\bm{p}}_{P}^{g}=\hat{\bm{p}}_{P}^{g}(\alpha_{\theta})$,
i.e., a continuum of initial guesses. Therefore, in order to obtain
a (small) finite set of initial guesses we apply the following steps:
\begin{enumerate}
\item define a uniform grid of $N_{\theta}$ values on $\alpha_{\theta}$,
constrained in the interval $[\alpha_{\theta,\min},\alpha_{\theta,\max}]$%
\footnote{The values $\{\alpha_{\theta,\min},\alpha_{\theta,\max}\}$ are obtained
automatically, once a reasonable range is given for parameters concurring
with the definition of $\alpha_{\theta}$ (e.g. $\sigma_{\theta}$);
an example will be given in Section \ref{sec:Simulation-Results}.%
} and, after denoting the $m$-th value as $\alpha_{\theta}[m]=\alpha_{\theta,\min}+\frac{m-1}{N_{\theta}-1}(\alpha_{\theta,\max}-\alpha_{\theta,\min})$,
$m\in\mathcal{S}_{\theta}\triangleq\{1,\ldots,N_{\theta}\}$, we compute
$\hat{\bm{p}}_{P}^{g}(\alpha_{\theta}[m])$, $g\in\{1,2\}$, $m\in\mathcal{S}_{\theta}$; 
\item split up the set $\hat{\bm{p}}_{P}^{g}(\alpha_{\theta}[m])$, $m\in\mathcal{S}_{\theta},g\in\{-1,1\}$
into three subsets, corresponding to zones ($a$), ($b$) and ($c$),
defined as follows%
\footnote{Note that the definition of the zones is arbitrary.%
}: ($i$) we split $\hat{\bm{p}}_{P}^{g}(\alpha_{\theta}[m])$ into
two sets $\hat{\bm{p}}_{P}^{g=1}(\alpha_{\theta}[m])$ and $\hat{\bm{p}}_{P}^{g=-1}(\alpha_{\theta}[m])$;
$(ii)$ by defining $\Gamma_{m,g}\triangleq\arctan_{2}\left(\hat{\bm{p}}_{P}^{g}(t_{n},\alpha_{\theta}[m])-\hat{\bm{p}}_{P}^{g}(t_{1},\alpha_{\theta}[m])\right)$
we seek for $\{m^{*},g^{*}\}\ni\mathrm{sign}(\Gamma_{m^{*},g^{*}})\neq\mathrm{sign}(\Gamma_{(m^{*}+1),g^{*}})$;
($iii$) we split the set corresponding to $g^{*}$ as $\hat{\bm{p}}_{P}^{g^{*}}(\alpha_{\theta}[1:m^{*}])$
and $\hat{\bm{p}}_{P}^{g^{*}}(\alpha_{\theta}[(m^{*}+1):N_{\theta}])$;
\item choose, from each defined subset, the guess corresponding to the maximum
of Eq. (\ref{eq:final_obj_function}), and denote the obtained triple
as $\{\hat{\bm{p}}_{P}^{A},\hat{\bm{p}}_{P}^{B},\hat{\bm{p}}_{P}^{C}\}$,
which is fed (possibly in parallel) to a local optimization routine.
\end{enumerate}
A good choice of $N_{\theta}$ can be obtained as follows. We first
observe that the worst-case error in grid sampling of $\alpha_{\theta}$
is given by $\Delta\varepsilon_{\theta}=\frac{(\alpha_{\theta,max}-\alpha_{\theta,min})}{2\cdot(N_{\theta}-1)}$;
such error is negligible in Eq. (\ref{eq:sec_approx_ri}) when $\breve{\alpha}_{\theta}\gg\Delta\varepsilon_{\theta}$,
thus leading to $N_{\theta}\gg\frac{\alpha_{\theta,max}-\alpha_{\theta,min}}{2\breve{\alpha}_{\theta}}+1$.
However, since $\breve{\alpha}_{\theta}$ is not known, we can consider
the conservative inequality $\alpha_{\theta,min}\gg\Delta\varepsilon_{\theta}$,
which leads to $N_{\theta}\gg\frac{1}{2}\frac{\alpha_{\theta,min}+\alpha_{\theta,max}}{\alpha_{\theta,min}}$.

\section{Simulation Results \label{sec:Simulation-Results}}

\begin{table}
\centering{}\caption{Parameters known at the platform side.\label{tab:Platform_side_parameters}}
\begin{tabular}{c|c|c}
\hline 
Parameter  & Value & Unit\tabularnewline
\hline 
\noalign{\vskip\doublerulesep}
\hline 
$t_{n}-t_{1}$ & $800$ & $\unit{s}$\tabularnewline
\hline 
$t_{i}-t_{i-1}$, $i\in\mathcal{I}\backslash\{1\}$ & $4$ & $\unit{s}$\tabularnewline
\hline 
\noalign{\vskip\doublerulesep}
$\bm{\breve{x}}_{P}^{s}$ in scenario ($i$) & $\left[\begin{array}{ccccc}
10^{4} & 2\cdot10^{4} & 7.1 & \frac{3}{4}\pi & \frac{\pi}{4}\end{array}\right]^{t}$ & $\left[\begin{array}{ccccc}
\unit{m} & \unit{m} & \unitfrac{m}{s} & \unit{rad} & \unit{rad}\end{array}\right]^{t}$\tabularnewline[\doublerulesep]
\hline 
\noalign{\vskip\doublerulesep}
$\bm{\breve{x}}_{P}^{s}$ in scenario ($ii$) & $\left[\begin{array}{ccccc}
10^{4} & 2\cdot10^{4} & 7.1 & -\frac{\pi}{4} & \frac{\pi}{4}\end{array}\right]^{t}$ & $\left[\begin{array}{ccccc}
\unit{m} & \unit{m} & \unitfrac{m}{s} & \unit{rad} & \unit{rad}\end{array}\right]^{t}$\tabularnewline[\doublerulesep]
\hline 
$q_{2}(\lambda v_{g},P_{D},g)$ & $q_{2}(0.3,0.9,5)=0.814$ & $\unit{dimensionless}$\tabularnewline
\hline 
$\sigma_{\theta}$ & $1$ ($0.0175$) & $\unit{deg.}$ ($\unit{rad}$)\tabularnewline
\hline 
$\breve{\alpha}_{\theta}=\nicefrac{q_{2}}{\sigma_{\theta}^{2}}$ & $2.6580\cdot10^{3}$ & $\unit{rad^{-2}}$\tabularnewline
\hline 
\end{tabular}
\end{table}

\begin{table}
\centering{}\caption{Parameters known at the target-friendly side.\label{tab:Target_side_parameters}}
\begin{tabular}{c|c|c}
\hline 
Parameter  & Value & Unit\tabularnewline
\hline 
\noalign{\vskip\doublerulesep}
\hline 
$t_{n}-t_{1}$ & $800$ & $\unit{s}$\tabularnewline
\hline 
$t_{i}-t_{i-1}$, $i\in\mathcal{I}\backslash\{1\}$ & $4$ & $\unit{s}$\tabularnewline
\hline 
$\bm{\hat{x}}_{T}$ & $[\begin{array}{cccc}
15\times10^{3} & 35\times10^{3} & -10 & 5\end{array}]^{t}$ & $[\begin{array}{cccc}
\unit{m} & \unit{m} & \unitfrac{m}{s} & \unitfrac{m}{s}\end{array}]^{t}$\tabularnewline
\hline 
$k$ & $101$ & $\unit{dimensionless}$\tabularnewline
\hline 
$q_{2}${*} & $\in[0.652,0.982]$ & $\unit{dimensionless}$\tabularnewline
\hline 
$\sigma_{\theta}${*} & $\in[1,2]$ ($\in[0.0175,0.0350]$) & $\unit{deg.}$ ($\unit{rad}$)\tabularnewline
\hline 
$\alpha_{\theta}${*} & $\in[532.2449,3206.5]$ & $\unit{rad^{-2}}$\tabularnewline
\hline 
\end{tabular}
\end{table}

In this section we consider two scenarios, taken from \cite{Bar-Shalom2002},
to corroborate the theoretical results presented and show the performance
of the geometry-driven initial guess procedure. In both scenarios
($i$) and ($ii$) we assume the same $\bm{\hat{x}}_{T}$ (since the
absolute position is irrelevant, only the relative geometry), while
we consider two different $\bm{\breve{x}}_{P}^{s}$ (see Table \ref{tab:Platform_side_parameters}).
In Table \ref{tab:Platform_side_parameters} we report the list of
parameters known at the platform side; it is worth noting that $\breve{\alpha}_{\theta}$
is completely specified by $q_{2}$ (obtained as in \cite[Table II]{Jauffret1990})
and $\sigma_{\theta}$. 

In Table \ref{tab:Target_side_parameters} we report the list of parameters
known at the target-friendly side, after intercepting the ML-PDA estimates.
The interval $[\alpha_{\theta,\min},\alpha_{\theta,\max}]$ (needed
to obtain the initial guesses in the three zones) is obtained as follows.
We assume that the target-friendly entity possesses the coarse information
$\sigma_{\theta}\in[1,2]^{\circ}$ and $q_{2}\in[0.652,0.982]$ (under
the assumptions%
\footnote{Note that in Table \ref{tab:Target_side_parameters} the asterisk
indicates that those parameters are only needed to compute wise initial
guesses for the local optimization procedure, but not for the maximization
of Eq. (\ref{eq:final_obj_function}).%
} that $\lambda v_{g}\in[0.1,0.5],$ $P_{D}\in[0.8,1]$, $g=5$, cf.
\cite[Table II]{Jauffret1990}), thus leading to $\alpha_{\theta}\in[532.2449,3206.5]\,\unit{rad^{-2}}$. 

The guesses $\{\hat{\bm{p}}_{P}^{A},\hat{\bm{p}}_{P}^{B},\hat{\bm{p}}_{P}^{C}\}$,
obtained with the approach described in Sec. \ref{sec:Initial Guess Choice}
and setting $N_{\theta}=5$%
\footnote{It is worth remarking that in such a case $\left[\frac{\alpha_{\theta,max}-\alpha_{\theta,min}}{2\breve{\alpha}_{\theta}}+1\right]\approx$1.5,
thus $N_{\theta}=5$ \emph{for this particular scenario} is a fair
choice.%
}, are given as input to the \emph{Nelder\textendash{}Mead simplex
}method\emph{ }\cite{Lagarias1998},\emph{ }which seeks for a local
maximum of the objective function in Eq. (\ref{eq:final_obj_function})\emph{.}
This local optimization routine has been chosen because it is a \emph{derivative-free
}method, as opposed to Newtonian and quasi-Newtonian local optimization
routines. Such a choice avoids the evaluation of the Jacobian matrix
(and also of the Hessian matrix, in the Newtonian approaches), which
is composed of $45$ entries, thus requiring extensive computations. 

\begin{figure}
\centering{}\includegraphics[scale=0.6]{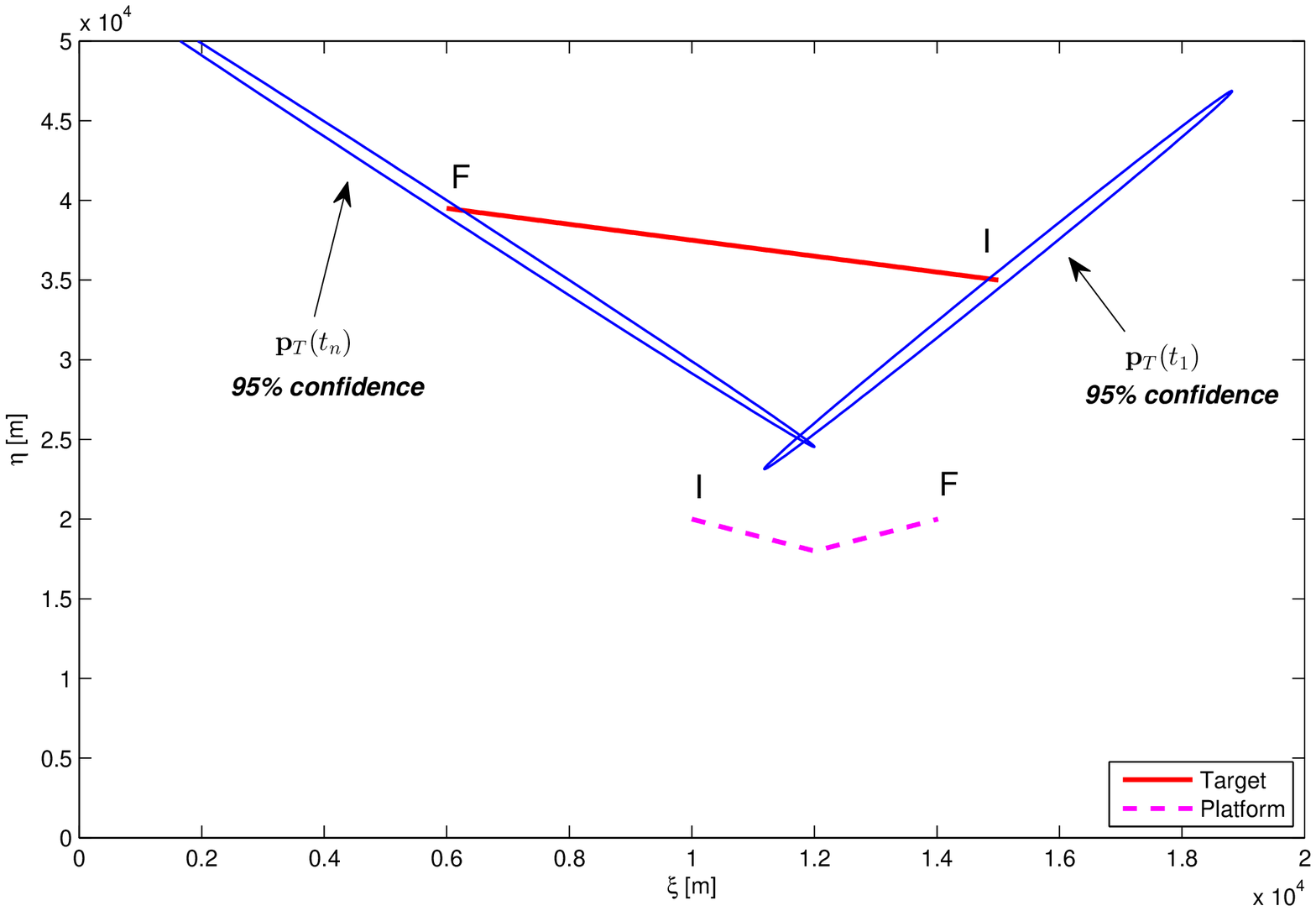}\caption{Platform and target trajectories considered for scenario ($i$).\label{fig:Platform-Target-ScenarioA}}
\end{figure}

\begin{figure}
\centering{}\includegraphics[scale=0.6]{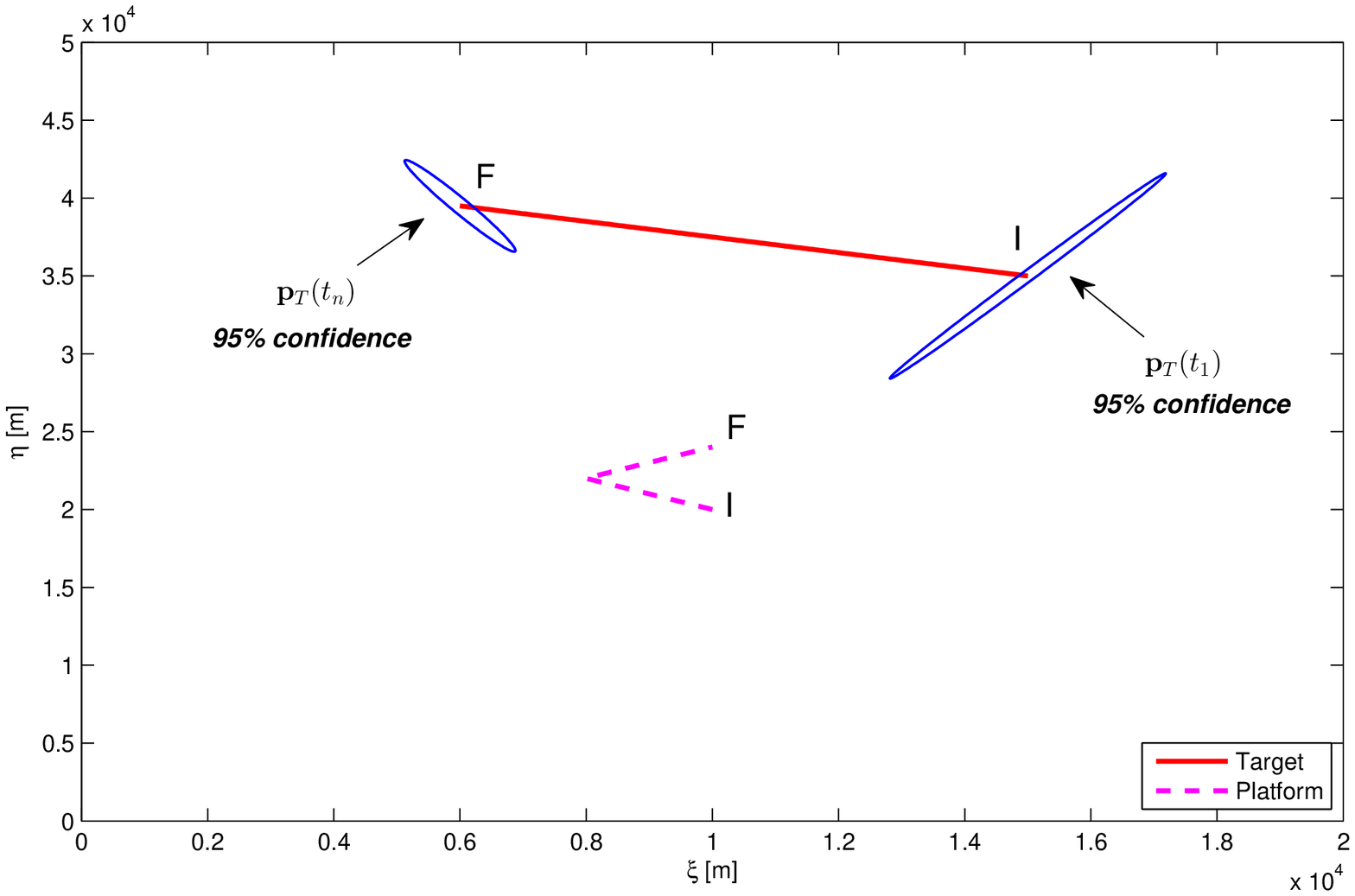}\caption{Platform and target trajectories considered for scenario ($ii$).\label{fig:Platform-Target-ScenarioB}}
\end{figure}

Fig. \ref{fig:Platform-Target-ScenarioA} presents a plot of the platform
and target trajectories for scenario ($i$), while Fig. \ref{fig:Platform-Target-ScenarioB}
shows them for scenario ($ii$). It is worth noting that scenario
($i$) represents a low-observability case, while scenario ($ii$)
has good observability, as shown through the $95\%$ confidence ellipses
of $\bm{\hat{p}}_{T}(t_{1})$ and $\bm{\hat{p}}_{T}(t_{n})$ in Figs.
\ref{fig:Platform-Target-ScenarioA} and \ref{fig:Platform-Target-ScenarioB}. 

\begin{figure}
\centering{}\includegraphics[scale=0.6]{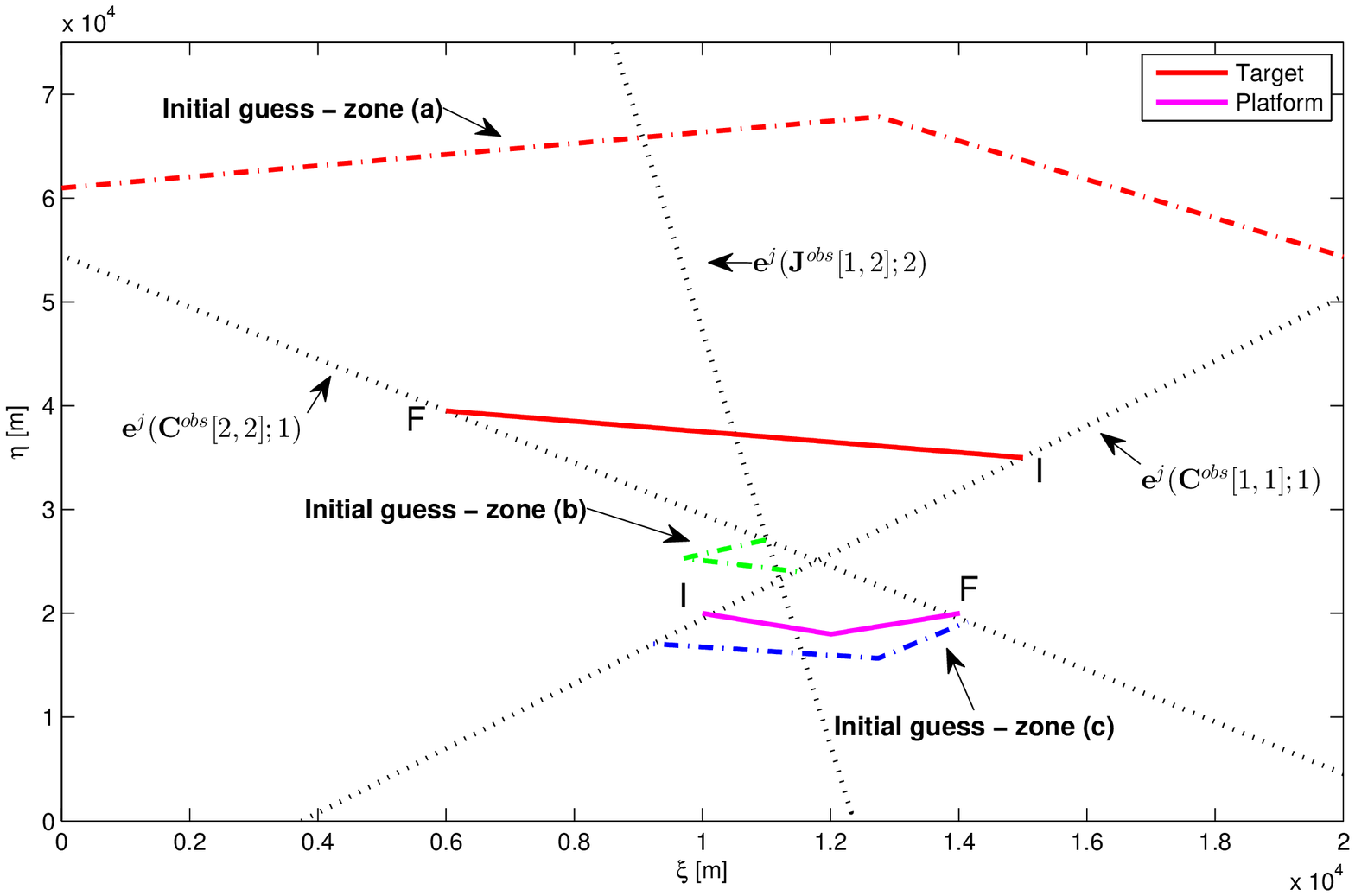}\caption{Geometry-driven initial guess procedure for platform trajectory in
scenario ($i$).\label{fig:Grid_search_ScenarioA}}
\end{figure}

\begin{figure}
\centering{}\includegraphics[scale=0.6]{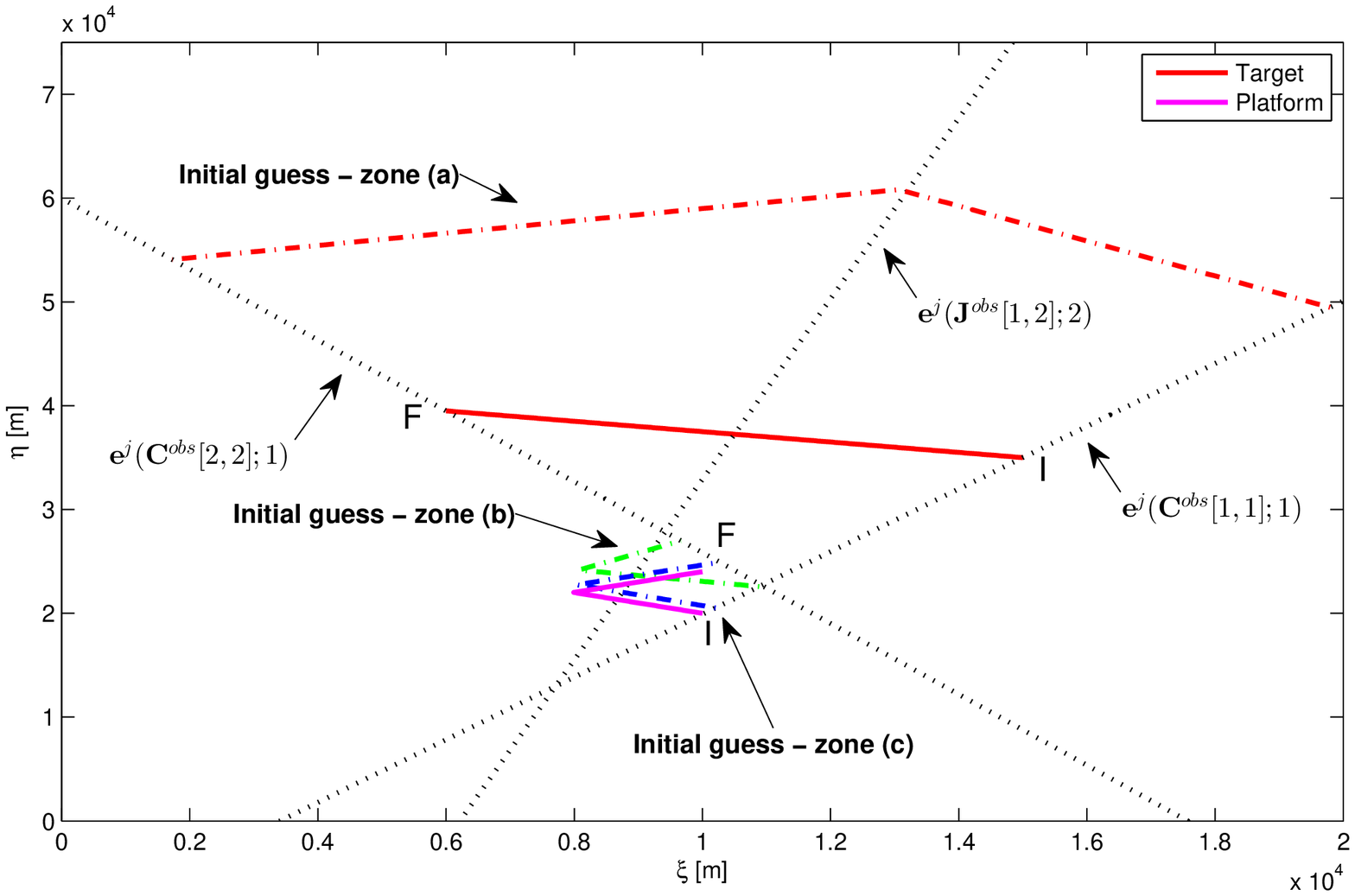}\caption{Geometry-driven initial guess procedure for platform trajectory in
scenario ($ii$).\label{fig:Grid_search_ScenarioB}}
\end{figure}

In Fig. \ref{fig:Grid_search_ScenarioA} (resp. Fig. \ref{fig:Grid_search_ScenarioB})
we show the initial estimates corresponding to zones ($a$), ($b$)
and ($c$) (defined in step $2$) of Subsec. \ref{sub:Remarks_alpha_theta})
of the platform trajectory (\ref{eq:final_obj_function}). It is apparent
that the initial estimate corresponding to the \emph{true zone }(i.e.
the zone where the true platform trajectory is) is near the true platform
trajectory in both cases; incidentally a degree of similarity is also
present in the estimate corresponding to a different zone in scenario
($ii$) (zone ($b$), cf. Fig. \ref{fig:Grid_search_ScenarioB}).
Nonetheless, in both cases the procedure produces a very good initial
estimate corresponding to the zone where $\bm{\breve{x}}_{P}^{s}$
belongs, which bodes well for a local optimization routine. 

\begin{figure}
\centering{}\includegraphics[scale=0.6]{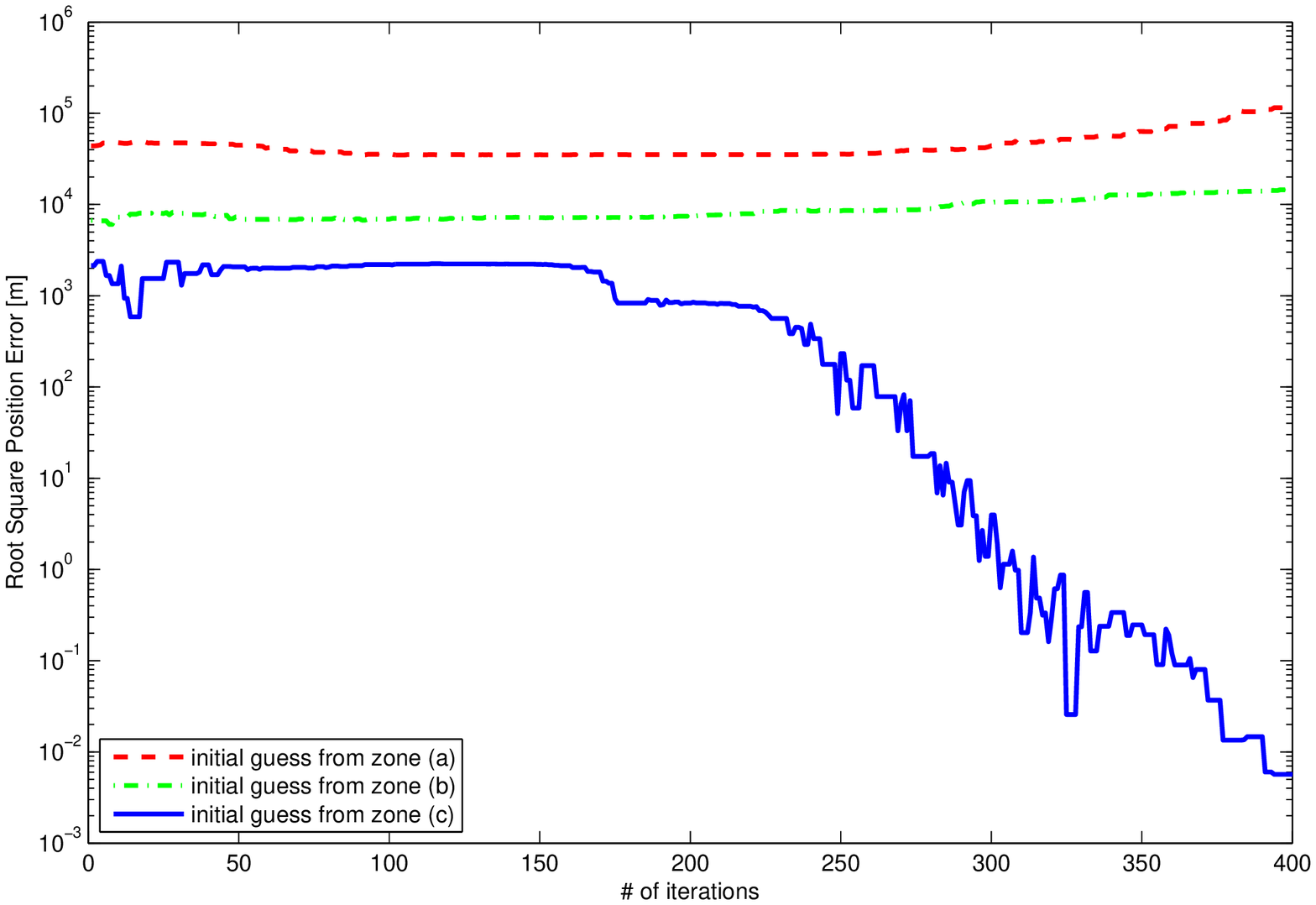}\caption{RSPE as a function of the number of iterations for scenario ($i$).\label{fig:PRSE-ScenarioA}}
\end{figure}

\begin{figure}
\centering{}\includegraphics[scale=0.6]{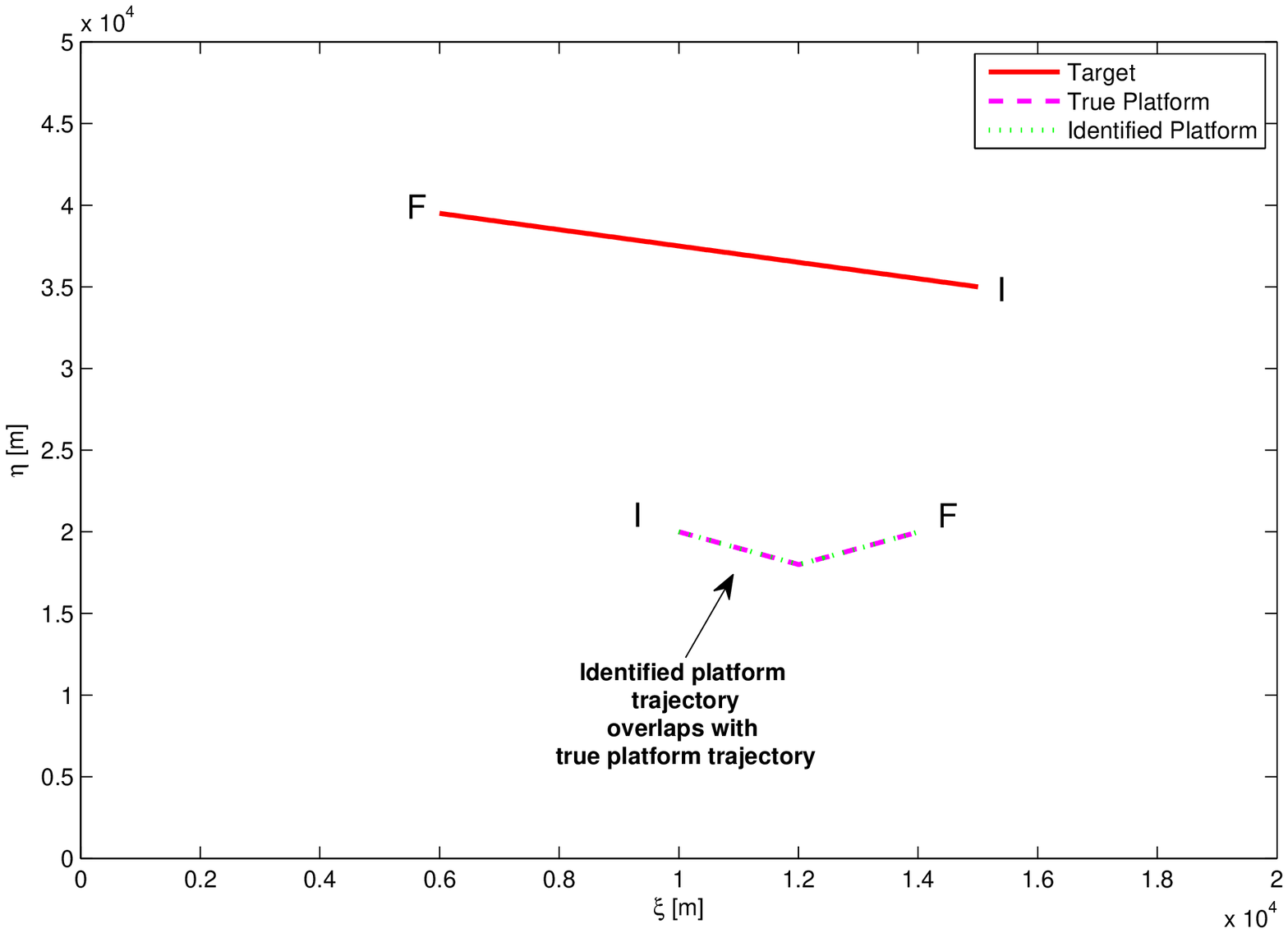}\caption{Identification results for scenario ($i$).\label{fig:FinalEstimate_ScenarioA}}
\end{figure}

\begin{figure}
\centering{}\includegraphics[scale=0.6]{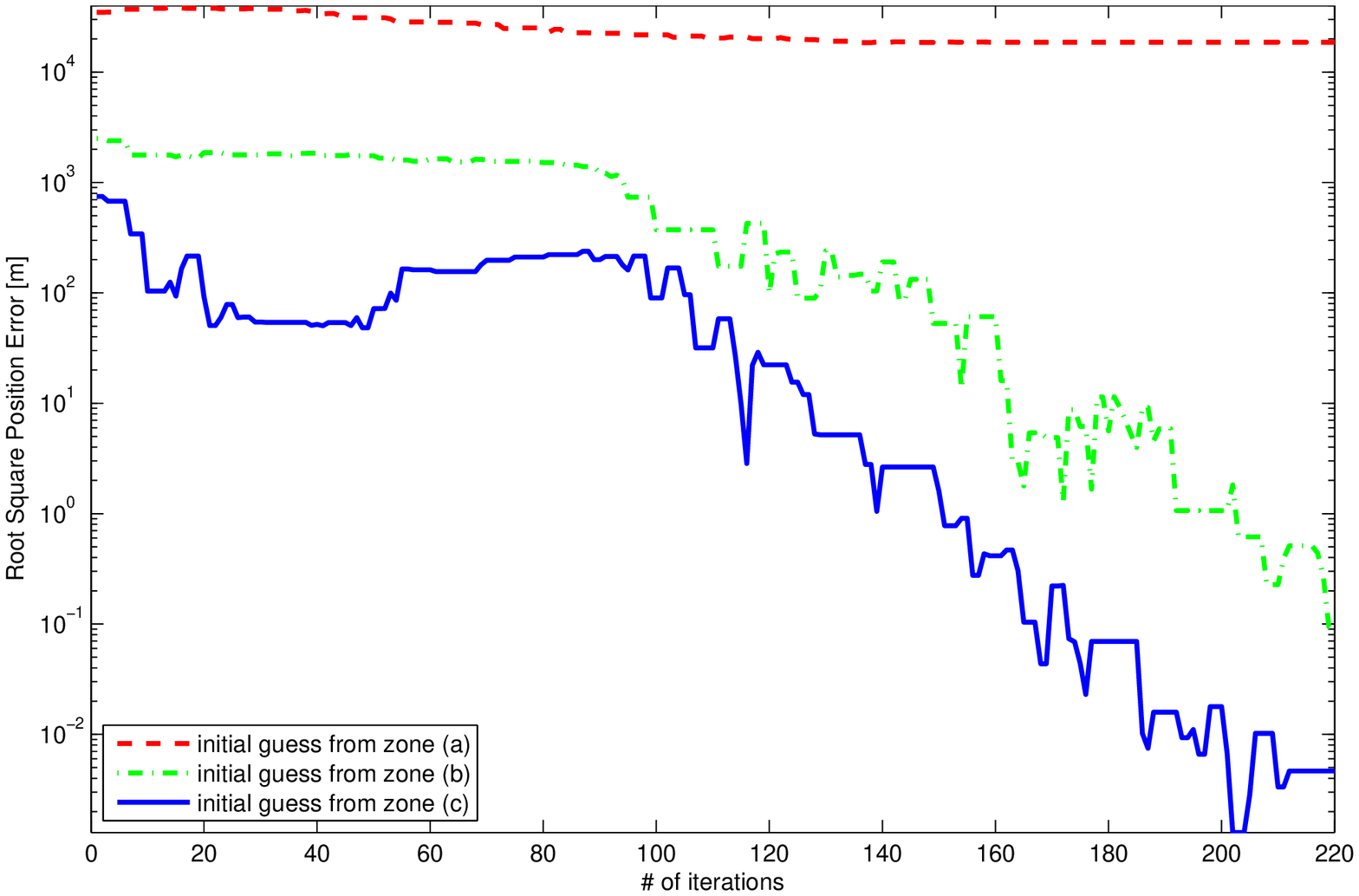}\caption{RSPE as a function of the number of iterations for scenario ($ii$).\label{fig:PRSE-ScenarioB}}
\end{figure}

\begin{figure}
\centering{}\includegraphics[scale=0.6]{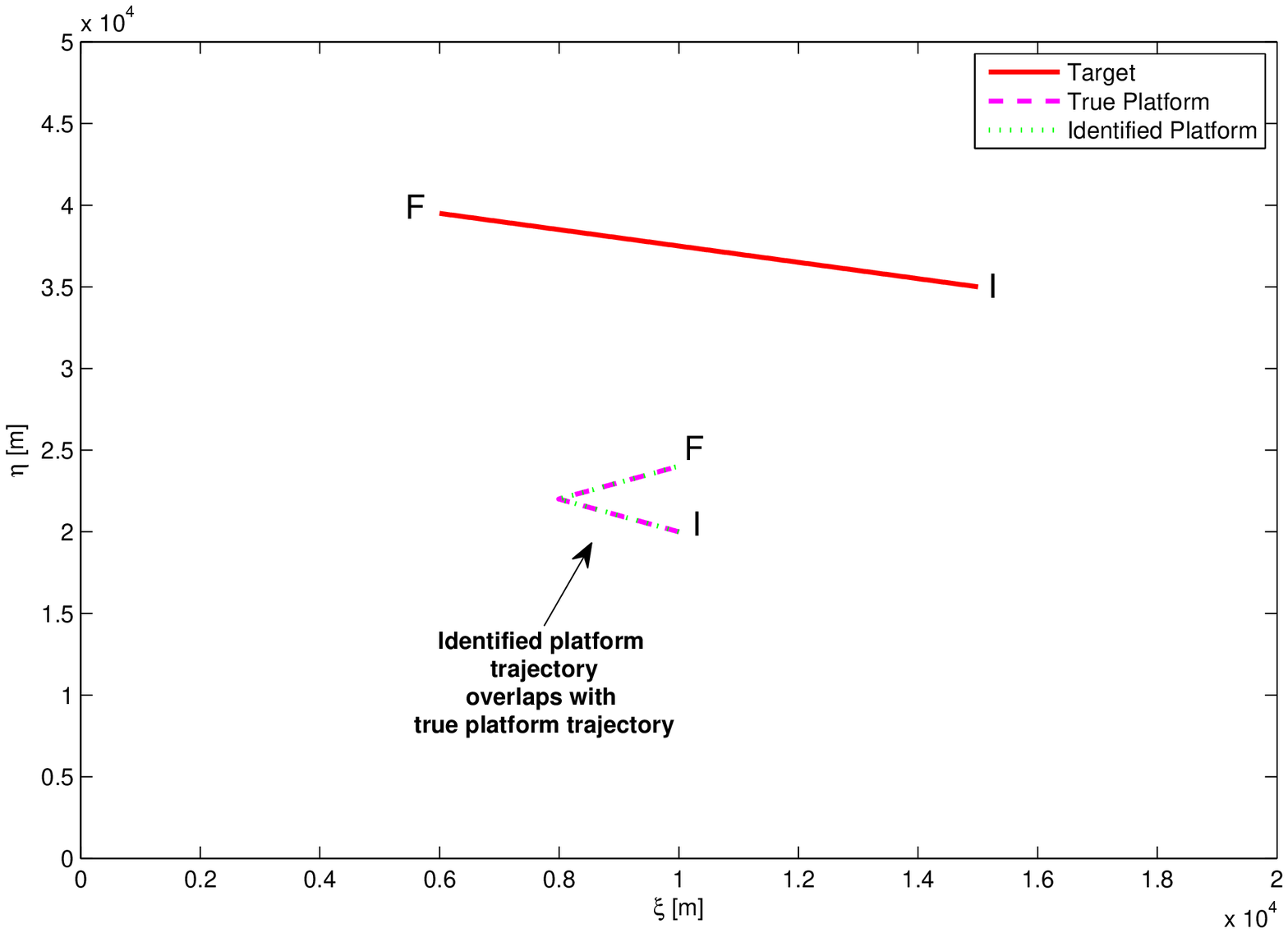}\caption{Identification results for scenario ($ii$). \label{fig:FinalEstimate_ScenarioB}}
\end{figure}

The convergence properties of the algorithm are illustrated in Figs.
\ref{fig:PRSE-ScenarioA} and \ref{fig:PRSE-ScenarioB} for the three
different initial estimates, and the performance is analyzed in terms
of the time-averaged root-square position-error (RSPE) defined as
\begin{equation}
\mathrm{RSPE}(m)\triangleq\frac{1}{n}\sum_{i=1}^{n}\left\Vert \bm{\hat{p}}_{P}^{m}(t_{i})-\bm{\breve{p}}_{P}(t_{i})\right\Vert _{2}\label{eq:RPSE}
\end{equation}
with $\bm{\hat{p}}_{P}^{m}(\cdot)$ denoting the output of the local
optimization routine after $m$ iterations. It is apparent that there
is no monotonic decrease of the RSPE, since the maximization of the
objection function is conducted w.r.t. the vector $\bm{x}_{P}^{s}$;
however convergence is observed with an acceptable number of iterations
(recall that there is no need to compute the Jacobian at each iteration
and so each iteration is very light from a computational point of
view). In Fig. \ref{fig:FinalEstimate_ScenarioA} (resp. Fig. \ref{fig:FinalEstimate_ScenarioB})
we finally show the platform corresponding to the maximum of the three
outputs obtained with the Nelder-Mead method and the three different
initial estimates. It is apparent that in both the scenarios the true
platform trajectory is identified exactly (recall that this is a deterministic
problem); such results confirm our conjecture on the uniqueness of
the Eq. (\ref{eq:nonlinear_system}).



\subsection*{Turning time $t_{k}$ sensitivity analysis }

\begin{figure}
\centering{}\includegraphics[scale=0.6]{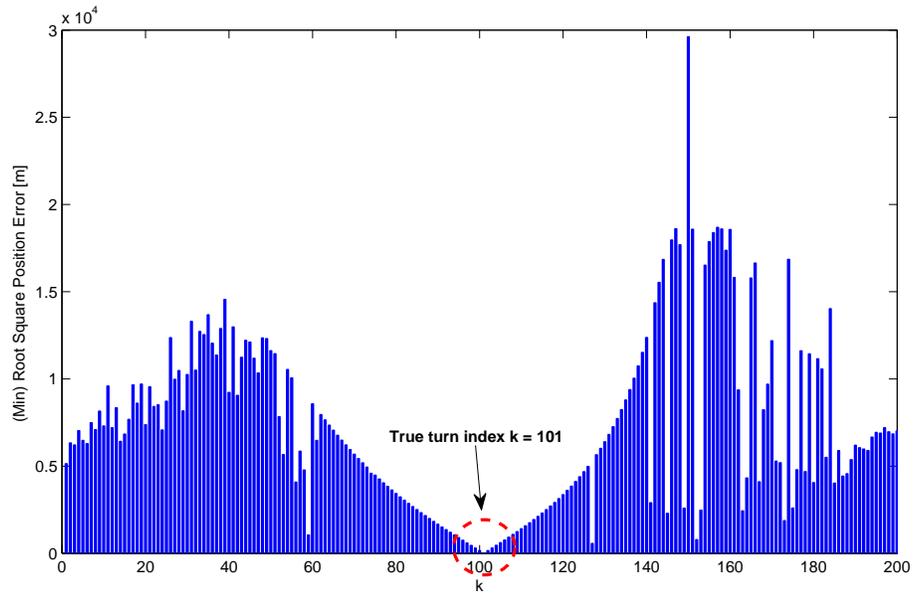}\caption{Sensitivity analysis in scenario ($i$).\label{fig:Sensitivity-scenario1}}
\end{figure}

\begin{figure}
\centering{}\includegraphics[scale=0.6]{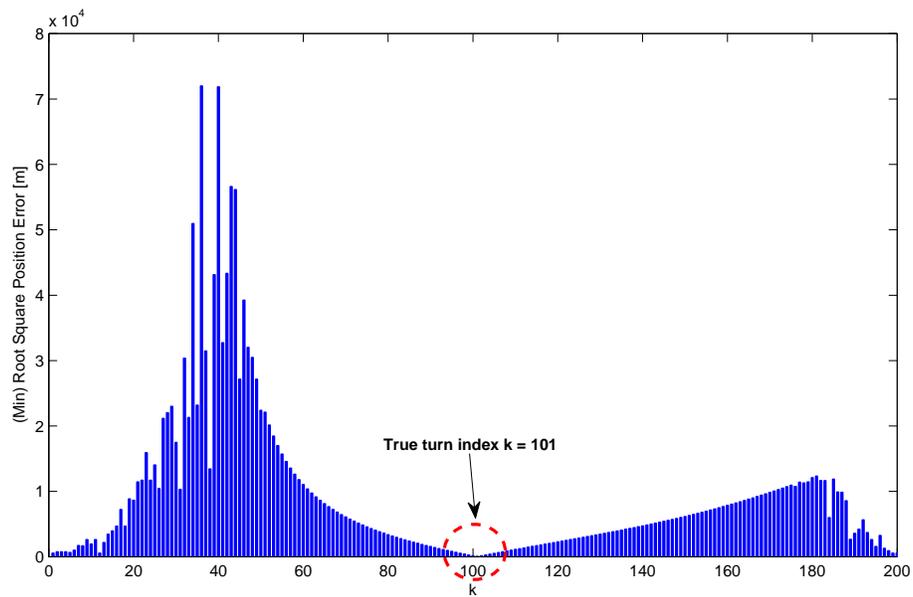}\caption{Sensitivity analysis in scenario ($ii$).\label{fig:Sensitivity-scenario2}}
\end{figure}

In this paragraph we will remove the assumption that the turning time
$t_{k}$ is known at the target-friendly side and we will show the
effects of a grid search for the turning time index, denoted as $k$,
for both scenarios ($i$) and ($ii$). Figs. \ref{fig:Sensitivity-scenario1}
and \ref{fig:Sensitivity-scenario2} show the RSPE (obtained as the
minimum along the three zones) as a function of the assumed index
$k$. It is apparent that the true platform trajectory is still identifiable
in both scenarios; however the RSPE w.r.t. to the index $k$ is not
a unimodal (discrete) function and therefore no ``naïve'' golden-search
method can be applied to identify the platform; rather a parallel
approach is needed.

Finally it is worth noting that the overall complexity $\Xi$ of the
proposed approach is given by:
\begin{eqnarray}
\Xi & \triangleq & \sum_{k=1}^{n}\omega_{a}(k)+\omega_{b}(k)+\omega_{c}(k)\\
 & = & \bar{\omega}\sum_{k=1}^{n}\chi_{a}(k)+\chi_{b}(k)+\chi_{c}(k)\\
\omega_{i}(k) & \triangleq & \bar{\omega}\cdot\chi_{i}(k)
\end{eqnarray}
where $\omega_{i}(k)$ denotes the complexity of the Nelder-Mead method
with initial estimate belonging to zone $i$ and assumed index $k$.
Note that $\omega_{i}(k)$ is simply given by the complexity of the
single iteration $\bar{\omega}$, multiplied by the number of iterations
$\chi_{i}(k)$. It is worth noting that there is no convergence theory
to support an analysis providing an estimate for the number of iterations
required to to satisfy any reasonable accuracy constraint, given as
a stopping condition \cite{Singer1999}. Instead, regarding the complexity
of the single iteration, it has been proved in \cite{Singer1999}
that $\bar{\omega}$ has a complexity  which is only dependent on
the dimension of the search space, which is five-dimensional  in our
case (since $\bm{x}_{P}^{s}\in\mathbb{R}^{5}$).


\section{Conclusions \label{sec:Conclusions}}

In this paper we studied the problem of identifying the platform motion
from its ML-PDA estimation results on an observed target\textemdash{}the
estimation of the stealthy estimator. We have addressed only the common
``two-leg'' platform trajectory (see Figs. \ref{fig:Target Estimation},
\ref{fig:Platform identification}, \ref{fig:System Model}, \ref{fig:Intersection-Eigenvectors}
etc.); a similar analysis could be performed to other platform trajectories
(such as a constant speed turn). We demonstrated that even a general
``two-leg'' platform motion model can lead to ambiguity in the identification;
however, imposition of the constraint of constant speed motion ensures
observability of the platform in most of the scenarios. We modelled
the problem as a Frobenius norm minimization and we found a convenient
objective function, exploiting the FIM elements and independent on
the platform measurement-related parameters, namely $\alpha_{\theta}$.
Also, we devised a procedure for the choice of a very small set of
initial estimates as input for the numerical optimization procedure,
on the basis of theoretical considerations on the geometry of the
FIM. Finally, we corroborated the theoretical findings and we have
shown the effectiveness of the approach for the choice of the initial
estimates, through simulation results. Future research will tackle
incomplete (and noisy) intercepted information and different platform
motion models. Note that although we have taken the ML-PDA as the
underlying algorithm this is only in attempt at generality. In fact,
the results presented here apply also to situations with any degree
of measurement origin uncertainty, including of course the situation
of a deterministic target observed via ``clean'' measurements without
false alarms or missed detections.

\section{Acknowledgement}

The authors would like to thank the editor and the anonymous reviewers
for their valuable comments and especially for one encouraging us
to address one particular key issue of observability.

\appendices{}

\section{Proof of Proposition \ref{prop:unobservability_statement}\label{sec:Appendix-unobservability-statement}}

To prove this proposition let us assume that there exists a solution
$\{\bm{x}_{P}^{*},\alpha_{\theta}^{*}\}$ such that $\bm{J}(\bm{\hat{x}}_{T},\bm{x}_{P}^{*},\alpha_{\theta}^{*})=\bm{J}^{obs}$.
Now let us consider the subspace $\{\bm{x}_{P}^{'},\alpha_{\theta}^{'}\}$
defined as 
\begin{eqnarray}
\{\bm{x}_{P}^{'},\alpha_{\theta}^{'}\} & = & \left\{ \beta\bm{x}_{P}^{*}+(1-\beta)\bm{\bar{x}}_{E},\beta^{2}\alpha_{\theta}^{*}\right\} ,\quad\beta\in\mathbb{R}\label{eq:unobservable_subspace_def_Appendix}
\end{eqnarray}
where $\bm{\bar{x}}_{E}\triangleq\left[\begin{array}{ccc}
\bm{\hat{p}}_{T}(t_{1})^{t} & \bm{\hat{v}}_{T}^{t} & \bm{\hat{v}}_{T}^{t}\end{array}\right]^{t}$. The subspace contains the set of platform trajectories whose velocity
and position vectors are linear combinations of the ones of the platform
trajectory defined by $\bm{x}_{P}^{*}$ and the estimated target trajectory. 

In this case the corresponding $\theta_{i}(\bm{\hat{x}}_{T},\bm{x}_{P}^{'})$
and $r_{i}(\bm{\hat{x}}_{T},\bm{x}_{P}^{'})$ have the explicit expressions
\begin{align}
\theta_{i}(\bm{\hat{x}}_{T},\bm{x}_{P}^{'}) & =\arctan\left(\frac{\hat{\xi}_{T}(t_{i})-\left(\beta\xi_{P}(t_{i})+(1-\beta)\hat{\xi}_{T}(t_{i})\right)}{\hat{\eta}_{T}(t_{i})-\left(\beta\eta_{P}(t_{i})+(1-\beta)\hat{\eta}_{T}(t_{i})\right)}\right)\nonumber \\
 & =\theta_{i}(\bm{\hat{x}}_{T},\bm{x}_{P}^{*})\label{eq:bearing_unobservable}
\end{align}
\begin{eqnarray}
r_{i}(\bm{\hat{x}}_{T},\bm{x}_{P}^{'}) & = & \sqrt{\left(\hat{\xi}_{T}(t_{i})-\left(\beta\xi_{P}(t_{i})+(1-\beta)\hat{\xi}_{T}(t_{i})\right)\right)^{2}+\left(\hat{\eta}_{T}(t_{i})-\left(\beta\eta_{P}(t_{i})+(1-\beta)\hat{\eta}_{T}(t_{i})\right)\right)^{2}}\nonumber \\
 & = & \beta r_{i}(\bm{\hat{x}}_{T},\bm{x}_{P}^{*})\label{eq:range_unobservable}
\end{eqnarray}

By plugging Eqs. (\ref{eq:bearing_unobservable}) and (\ref{eq:range_unobservable})
into Eq. (\ref{eq:gradient_vector_for_FIM}) we obtain the equality
\begin{equation}
\bm{\nabla}_{\bm{x}_{T}}(\theta_{i}(\bm{\hat{x}}_{T},\bm{x}_{P}^{'}))=\frac{1}{\beta}\bm{\nabla}_{\bm{x}_{T}}(\theta_{i}(\bm{\hat{x}}_{T},\bm{x}_{P}^{*}))\label{eq:gradient_equality_appendixA}
\end{equation}
 Using Eq. (\ref{eq:gradient_equality_appendixA}) we can express
$\bm{J}(\bm{\hat{x}}_{T},\bm{x}_{P}^{'},\alpha_{\theta}^{'})$ as
\begin{eqnarray}
\bm{J}(\bm{\hat{x}}_{T},\bm{x}_{P}^{'},\alpha_{\theta}^{'}) & = & \left(\beta^{2}\alpha_{\theta}^{*}\right)\sum_{i=0}^{n}\frac{1}{\beta}\bm{\nabla}_{\bm{x}_{T}}(\theta_{i}(\bm{\hat{x}}_{T},\bm{x}_{P}^{*}))\frac{1}{\beta}\bm{\nabla}_{\bm{x}_{T}}^{t}(\theta_{i}(\bm{\hat{x}}_{T},\bm{x}_{P}^{*}))\\
 & = & \bm{J}(\bm{\hat{x}}_{T},\bm{x}_{P}^{*},\alpha_{\theta}^{*})
\end{eqnarray}
 Since $\{\bm{x}_{P}^{*},\alpha_{\theta}^{*}\}$ and $\{\bm{x}_{P}^{'},\alpha_{\theta}^{'}\}$
(\emph{independently} of $\beta$) lead both to $\bm{J}^{obs}$, $\bm{J}(\cdot)$
does not represent a \emph{one-to-one mapping, }which makes the platform-state
vector $\bm{x}_{P}$ unidentifiable.

\section{Proof of Lemma \ref{lem:Fisher_Matrix_9elements}\label{sec:Proof-of-9 elements FIM}}

We start by observing that $\bm{J}(\bm{\hat{x}}_{T},\bm{x}_{P}^{s},\alpha_{\theta})$
and $\bm{\nabla}_{\bm{x}_{T}}(\theta_{i}(\bm{\hat{x}}{}_{T},\bm{x}_{P}^{s}))$
can be expressed similarly as Eqs. (\ref{eq:FIM_definition}) and
(\ref{eq:gradient_vector_for_FIM}). Also, for notational convenience
let us define 
\begin{eqnarray}
\bm{y}_{i} & \triangleq & \frac{1}{r_{i}(\bm{\hat{x}}_{T},\bm{x}_{P}^{s})}\left[\begin{array}{cc}
\cos(\theta_{i}(\bm{\hat{x}}_{T},\bm{x}_{P}^{s})) & -\sin(\theta_{i}(\bm{\hat{x}}_{T},\bm{x}_{P}^{s}))\end{array}\right]^{t}
\end{eqnarray}
 and rewrite $\bm{\nabla}_{\bm{x}_{T}}(\theta_{i}(\bm{\hat{x}}{}_{T},\bm{x}_{P}^{s}))$
as 
\begin{equation}
\bm{\nabla}_{\bm{x}_{T}}(\theta_{i}(\bm{x}_{T},\bm{x}_{P}^{s}))=\left[\begin{array}{c}
(1-\alpha_{i})\bm{y}_{i}\\
\alpha_{i}\bm{y}_{i}
\end{array}\right]\label{eq:gradient_shorhand_notation}
\end{equation}
where $\alpha_{i}$ has been defined in Eq. (\ref{eq:alternative target motion model}).
Substituting Eq. (\ref{eq:gradient_shorhand_notation}) into the explicit
form of $\bm{J}(\bm{\hat{x}}_{T},\bm{x}_{P}^{s},\alpha_{\theta})$,
we get
\begin{eqnarray}
\bm{J}(\bm{\hat{x}}_{T},\bm{x}_{P}^{s},\alpha_{\theta}) & \triangleq & \left[\begin{array}{cc}
\bm{J}[1,1] & \bm{J}[1,2]\\
\bm{J}[2,1] & \bm{J}[2,2]
\end{array}\right]\label{eq:FIM-symmetric}\\
 & = & \alpha_{\theta}\left[\begin{array}{cc}
\sum_{i=1}^{n}(1-\alpha_{i})^{2}\bm{y}_{i}\bm{y}_{i}^{t} & \sum_{i=1}^{n}\alpha_{i}(1-\alpha_{i})\bm{y}_{i}\bm{y}_{i}^{t}\\
\sum_{i=1}^{n}\alpha_{i}(1-\alpha_{i})\bm{y}_{i}\bm{y}_{i}^{t} & \sum_{i=1}^{n}\alpha_{i}{}^{2}\bm{y}_{i}\bm{y}_{i}^{t}
\end{array}\right]\label{eq:FIM-symmetric-block-expression}
\end{eqnarray}
where we have elucidated the block-decomposition arising from Eq.
(\ref{eq:gradient_shorhand_notation}) into $[2\times2]$ matrices
$\bm{J}[\ell,m]$, $\ell,m\in\{1,2\}$. Since we have that $\bm{J}[1,2]=\bm{J}[2,1]$,
only three matrices contain non-repeated entries (i.e., the $4$ entries
of $\bm{J}[1,2]$ or $\bm{J}[2,1]$ can be neglected). Also, since
$\bm{J}[1,1]$, $\bm{J}[2,1]$ and $\bm{J}[2,2]$ are symmetric matrices,
there is a repeated entry in each of them, thus leading to $3$ other
dependent entries. Therefore the $16$ entries of the FIM actually
contain only $9$ independent elements. W.l.o.g. we consider here
(and throughout the paper), the following independent entries (we
drop the dependence w.r.t. $\bm{\hat{x}}_{T}$, $\bm{x}_{P}^{s}$
and $\alpha_{\theta}$) and we denote with $\mathcal{C}$ the corresponding
set of indices: 
\begin{equation}
\{J_{1,1},J_{2,2},J_{3,3},J_{4,4},J_{1,2},J_{1,3},J_{1,4},J_{2,4},J_{3,4}\}\label{eq:Indepedent_Elements_FIM}
\end{equation}

\section{Proof of Proposition \ref{prop:weighted_squared_matrix_FIM} \label{sec:Appendix-weighted squared distance}}

We have shown, through Lemma \ref{lem:Fisher_Matrix_9elements}, that
$\bm{J}(\bm{x}_{P}^{s},\alpha_{\theta})$ has only $9$ independent
entries. Also, the dependent entries are simple repetitions of the
corresponding independent entries, due to particular symmetry structure
of the FIM (cfr. Eq. (\ref{eq:FIM-symmetric})). 

The same argument extends to $\bm{D}(\bm{x}_{P}^{s},\alpha_{\theta})\triangleq\left(\bm{J}^{obs}-\bm{J}(\bm{x}_{P}^{s},\alpha_{\theta})\right)$
and its entry-wise squared version $\bm{\bar{D}}(\bm{x}_{P}^{s},\alpha_{\theta})$
(i.e. $\bar{D}_{\ell,m}(\bm{x}_{P}^{s},\alpha_{\theta})=D_{\ell,m}^{2}(\bm{x}_{P}^{s},\alpha_{\theta})$),
\emph{iff} $\bm{J}^{obs}$ retains the same property of symmetry;
this is accomplished if $\bm{J}^{obs}$ is a noise-free observed FIM,
that is $\bm{J}^{obs}=\bm{J}(\bm{\breve{x}}_{P}^{s},\breve{\alpha}_{\theta})$. 

By construction, the following equality holds
\begin{equation}
\left\Vert \bm{J}^{obs}-\bm{J}(\bm{x}_{P}^{s},\alpha_{\theta})\right\Vert _{F}^{2}=\sum_{\ell=1}^{4}\sum_{m=1}^{4}\bar{D}_{\ell,m}(\bm{x}_{P}^{s},\alpha_{\theta})\label{eq:sum_elements_squared_matrix}
\end{equation}
This sum can be efficiently evaluated by considering only $(\ell,m)\in\mathcal{C}$,
with $\mathcal{C}$ being the set of the independent entries according
to Eq. (\ref{eq:Indepedent_Elements_FIM}), and weighting them by
the number of times they are repeated in the matrix $\bar{\bm{D}}$.
Thus the l.h.s. of Eq. (\ref{eq:sum_elements_squared_matrix}) can
be rewritten as
\begin{equation}
\sum\sum_{(\ell,m)\in\mathcal{C}}W_{\ell,m}\bar{D}_{\ell,m}(\bm{x}_{P}^{s},\alpha_{\theta})=\sum\sum_{(\ell,m)\in\mathcal{C}}W_{\ell,m}\left(J_{\ell,m}^{obs}-J_{\ell,m}(\bm{x}_{P}^{s},\alpha_{\theta})\right)^{2}\label{eq:App_weighted_sum_of_squares}
\end{equation}
By stacking the elements $J_{\ell,m}^{obs}$ (resp. $J_{\ell,m}(\bm{x}_{P}^{s},\alpha_{\theta})$),
$(\ell,m)\in\mathcal{C}$, into the vector $\bm{j}^{obs}$ (resp.
$\bm{j}(\bm{x}_{P}^{s},\alpha_{\theta})$), and defining the diagonal
matrix $\bm{W}$ with elements $W_{\ell,m}$, we obtain the weighted
form of Eq. (\ref{eq:weighted_nonlinear_LS}). Finally, it is straightforward
to show that $\bm{W}$ has the expression
\begin{equation}
\bm{W}=\mathrm{diag}\left(\left[\begin{array}{ccccccccc}
1 & 1 & 1 & 1 & 2 & 2 & 4 & 2 & 2\end{array}\right]^{t}\right)\label{eq:Weight_Matrix_NLS}
\end{equation}
where the weight equal to $4$ accounts for the repeated elements
along the minor (secondary) diagonal, while the weights equal to $2$
account for the remaining symmetric entries into each block matrix
$\bm{J}[\ell,m]$, $\ell,m\in\{1,2\}$, and between the matrices $\bm{J}[1,2]$
and $\bm{J}[2,1]$.

\section{Proof of Proposition \ref{prop:new_objective_function}\label{sec:Appendix_new_objective_function_demonstration}}

Let us rewrite Eq. (\ref{eq:weighted_nonlinear_LS_separated}) here
for convenience:
\begin{equation}
\mathcal{F}(\bm{x}_{P}^{s},\alpha_{\theta})=\left[\bm{j}^{obs}-\bm{j}_{u}(\bm{x}_{P}^{s})\alpha_{\theta}\right]^{t}\bm{W}\left[\bm{j}^{obs}-\bm{j}_{u}(\bm{x}_{P}^{s})\alpha_{\theta}\right]\label{eq:nonlinear_weighted_squared_distance_separated_app}
\end{equation}
For a fixed $\bm{x}_{P}^{s}$ it is easy to check that the observation
model in Eq. (\ref{eq:nonlinear_weighted_squared_distance_separated_app})
is linear in $\alpha_{\theta}$. Therefore given $\bm{x}_{P}^{s}$,
$\hat{\alpha}_{\theta}$ can be found as the solution of a standard
least squares problem \cite{Kay1993} as
\begin{equation}
\hat{\alpha}_{\theta}=\frac{\bm{j}_{u}(\bm{x}_{P}^{s})^{t}\bm{W}\bm{j}^{obs}}{\bm{j}_{u}(\bm{x}_{P}^{s})^{t}\bm{W}\bm{j}_{u}(\bm{x}_{P}^{s})}\label{eq:a_theta_lls}
\end{equation}
Then substituting Eq. (\ref{eq:a_theta_lls}) in Eq. (\ref{eq:nonlinear_weighted_squared_distance_separated_app})
we obtain the expression for $\mathcal{F}(\bm{x}_{P}^{s},\alpha_{\theta})$
\begin{equation}
\mathcal{F}(\bm{x}_{P}^{s},\hat{\alpha}_{\theta})=\left[\bm{j}^{obs}-\bm{j}_{u}(\bm{x}_{P}^{s})\frac{\bm{j}_{u}(\bm{x}_{P}^{s})^{t}\bm{W}\bm{j}^{obs}}{\bm{j}_{u}(\bm{x}_{P}^{s})^{t}\bm{W}\bm{j}_{u}(\bm{x}_{P}^{s})}\right]^{t}\bm{W}\left[\bm{j}^{obs}-\bm{j}_{u}(\bm{x}_{P}^{s})\frac{\bm{j}_{u}(\bm{x}_{P}^{s})^{t}\bm{W}\bm{j}^{obs}}{\bm{j}_{u}(\bm{x}_{P}^{s})^{t}\bm{W}\bm{j}_{u}(\bm{x}_{P}^{s})}\right]\label{eq:intermediate_new_objfunction1}
\end{equation}
Exploiting the \emph{orthogonality principle }of linear least squares
\cite{Kay1993}, Eq. (\ref{eq:intermediate_new_objfunction1}) reduces
to
\begin{equation}
\mathcal{F}(\bm{x}_{P}^{s},\hat{\alpha}_{\theta})=(\bm{j}^{obs})^{t}\left(\bm{W}-\frac{\bm{W}\bm{j}_{u}(\bm{x}_{P}^{s})\bm{j}_{u}(\bm{x}_{P}^{s})^{t}\bm{W}}{\bm{j}_{u}(\bm{x}_{P}^{s})^{t}\bm{W}\bm{j}_{u}(\bm{x}_{P}^{s})}\right)\bm{j}^{obs}\label{eq:eq:intermediate_new_objfunction2}
\end{equation}
Since $(\bm{j}^{obs})^{t}\bm{W}\bm{j}^{obs}$ in Eq. (\ref{eq:eq:intermediate_new_objfunction2})
does not depend on $\bm{x}_{P}^{s}$,it is irrelevant in the minimization.
Thus neglecting it and considering the opposite of the remaining term,
we can equivalently maximize:
\begin{equation}
\mathcal{G}(\bm{x}_{P}^{s})\triangleq(\bm{j}^{obs})^{t}\frac{\bm{W}\bm{j}_{u}(\bm{x}_{P}^{s})\bm{j}_{u}(\bm{x}_{P}^{s})^{t}\bm{W}}{\bm{j}_{u}(\bm{x}_{P}^{s})^{t}\bm{W}\bm{j}_{u}(\bm{x}_{P}^{s})}\bm{j}^{obs}\label{eq:final_objective_function_1}
\end{equation}
Finally, by defining $\bm{c}(\bm{x}_{P}^{s})\triangleq\frac{\bm{W}\bm{j}_{u}(\bm{x}_{P}^{s})}{\left[\bm{j}_{u}(\bm{x}_{P}^{s})^{t}\bm{W}\bm{j}_{u}(\bm{x}_{P}^{s})\right]^{\nicefrac{1}{2}}}$,
we can rewrite Eq. (\ref{eq:final_objective_function_1}) as
\begin{eqnarray}
\mathcal{G}(\bm{x}_{P}^{s}) & = & (\bm{j}^{obs})^{t}\bm{c}(\bm{x}_{P}^{s})\bm{c}(\bm{x}_{P}^{s})^{t}\bm{j}^{obs}\\
 & = & \left\langle \bm{j}^{obs},\bm{c}(\bm{x}_{P}^{s})\right\rangle ^{2}
\end{eqnarray}
which concludes the proof.

\section{Choice of $\{\hat{r}_{1},\hat{r}_{n}\}$ \label{sec: Appendix_ choice r_hat i}}

In this section we will derive the expressions in Eq. (\ref{eq:sec_approx_ri}).
For the sake of simplicity we will use the short-hand notations $r_{i}\triangleq r_{i}(\bm{\hat{x}}_{T},\bm{\breve{x}}_{P}^{s})$
and $\theta_{i}\triangleq\theta_{i}(\bm{\hat{x}}_{T},\bm{\breve{x}}_{P}^{s})$.
We start by considering the block decomposition of $\bm{J}^{obs}$,
through Eq. (\ref{eq:FIM-symmetric}). Let us focus in particular
on $\bm{J}^{obs}[1,1]$ and $\bm{J}^{obs}[2,2]$, i.e. the diagonal
blocks, whose explicit expressions are given by
\begin{eqnarray}
\bm{J}^{obs}[1,1] & = & \breve{\alpha}_{\theta}\sum_{i=1}^{n}\frac{(1-\alpha_{i})^{2}}{r_{i}^{2}}\underbrace{\left[\begin{array}{cc}
\cos^{2}\left(\theta_{i}\right) & -\nicefrac{1}{2}\sin\left(2\theta_{i}\right)\\
-\nicefrac{1}{2}\sin\left(2\theta_{i}\right) & \sin^{2}\left(\theta_{i}\right)
\end{array}\right]}_{\triangleq\bm{D}_{i}}\label{eq:J11-block}\\
\bm{J}^{obs}[2,2] & = & \breve{\alpha}_{\theta}\sum_{i=1}^{n}\frac{\alpha_{i}{}^{2}}{r_{i}^{2}}\underbrace{\left[\begin{array}{cc}
\cos^{2}\left(\theta_{i}\right) & -\nicefrac{1}{2}\sin\left(2\theta_{i}\right)\\
-\nicefrac{1}{2}\sin\left(2\theta_{i}\right) & \sin^{2}\left(\theta_{i}\right)
\end{array}\right]}_{\triangleq\bm{D}_{i}}\label{eq:J22-block}
\end{eqnarray}
note that an analogous expression holds for $\bm{J}^{obs}[1,2]$,
i.e. $\bm{J}^{obs}[1,2]=\breve{\alpha}_{\theta}\sum_{i=1}^{n}\frac{\alpha_{i}(1-\alpha_{i})}{r_{i}^{2}}\bm{D}_{i}$.
It can be readily shown, through Eqs. (\ref{eq:J11-block}) and (\ref{eq:J22-block}),
that $\mathrm{tr}\left(\bm{J}^{obs}[\ell,\ell]\right)$, $\ell\in\{1,2\}$,
are given by
\begin{eqnarray}
\mathrm{tr}\left(\bm{J}^{obs}[1,1]\right) & = & \breve{\alpha}_{\theta}\left[\frac{1}{r_{1}^{2}}+\sum_{i=2}^{n}\frac{(1-\alpha_{i})^{2}}{r_{i}^{2}}\right]\label{eq:gamma11}\\
\mathrm{tr}\left(\bm{J}^{obs}[2,2]\right) & = & \breve{\alpha}_{\theta}\left[\frac{1}{r_{n}^{2}}+\sum_{i=1}^{n-1}\frac{\alpha_{i}{}^{2}}{r_{i}^{2}}\right]\label{eq:gamma22}
\end{eqnarray}
 From inspection of Eqs. (\ref{eq:gamma11}) and (\ref{eq:gamma22}),
it is apparent that each trace is a weighted sum (scaled by $\breve{\alpha}_{\theta}$)
of $\frac{1}{r_{i}^{2}}$. Note that, by definition, the following
inequalities hold $\forall i\in\mathcal{I}$:
\begin{equation}
(1-\alpha_{i})^{2}<(1-\alpha_{i+1})^{2},\qquad\alpha_{i+1}{}^{2}<\alpha_{i}^{2}\label{eq:alpha_i_inequalities}
\end{equation}
 By exploiting them, we have that in $\mathrm{tr}\left(\bm{J}^{obs}[1,1]\right)$
(resp. $\mathrm{tr}\left(\bm{J}^{obs}[2,2]\right)$) the term $\frac{1}{r_{1}^{2}}$
(resp. $\frac{1}{r_{n}^{2}}$) receives the highest weight, while
the term $\frac{1}{r_{n}^{2}}$ (resp. $\frac{1}{r_{1}^{2}}$) contributes
to the sum with zero weight. To obtain a good approximation (and avoid
biased estimates) of $\{r_{1},r_{n}\}$ we first consider convex combination
counterparts of Eqs. (\ref{eq:gamma11}) and (\ref{eq:gamma22}),
since the positive weights in $\mathrm{tr}\left(\bm{J}^{obs}[1,1]\right)$
and $\mathrm{tr}\left(\bm{J}^{obs}[2,2]\right)$ \emph{do not satisfy
the normalization property} (i.e. $\sum_{i=1}^{n}(1-\alpha_{i})^{2}\neq1$
and $\sum_{i=1}^{n}\alpha_{i}{}^{2}\neq1$). The reasons are twofold:
($i$) the weights $\alpha_{i}$ and $(1-\alpha_{i})$ appear in squared
form in Eqs. (\ref{eq:gamma11}) and (\ref{eq:gamma22}); and ($ii$)
even $\alpha_{i}$ and $(1-\alpha_{i})$ do not sum to one%
\footnote{For example in the case of uniform sampling it holds $\sum_{i=1}^{n}\alpha_{i}=\sum_{i=1}^{n}(1-\alpha_{i})=\frac{n}{2}$;
thus sum grows proportionally with the sample rate. The growth of
$\sum_{i=1}^{n}\alpha_{i}$ and $\sum_{i=1}^{n}(1-\alpha_{i})$ with
the sample rate is present also under the more general non-uniform
sampling assumption, but in the latter case $\sum_{i=1}^{n}\alpha_{i}\neq\sum_{i=1}^{n}(1-\alpha_{i})\neq\frac{n}{2}$.%
}. For such a reason we normalize $\mathrm{tr}\left(\bm{J}^{obs}[1,1]\right)$
(resp. $\mathrm{tr}\left(\bm{J}^{obs}[1,1]\right)$) by $\sum_{i=1}^{n}(1-\alpha_{i})^{2}$
(resp. $\sum_{i=1}^{n}\alpha_{i}{}^{2}$). Then, we note that if 
\begin{eqnarray}
(a)\quad r_{i} & \approx & r_{1},\quad i\:\ni\:\alpha_{i}^{2}\ll1\nonumber \\
(b)\quad r_{j} & \approx & r_{n},\quad j\:\ni\:(1-\alpha_{j})^{2}\ll1\label{eq:new_range_assumption}
\end{eqnarray}
the term $\frac{\mathrm{tr}\left(\bm{J}^{obs}[1,1]\right)}{\sum_{i=1}^{n}(1-\alpha_{i})^{2}}$
(resp. $\frac{\mathrm{tr}\left(\bm{J}^{obs}[2,2]\right)}{\sum_{i=1}^{n}\alpha_{i}{}^{2}}$)
well approximates $\frac{\breve{\alpha}_{\theta}}{r_{1}^{2}}$ (resp.
$\frac{\breve{\alpha}_{\theta}}{r_{n}^{2}}$), thus leading to Eq.
(\ref{eq:sec_approx_ri}); therefore the value of $\mathrm{tr}\left(\bm{J}^{obs}[1,1]\right)$
(resp. $\mathrm{tr}\left(\bm{J}^{obs}[2,2]\right)$) will be high
when the platform trajectory is near to the target at the beginning
(resp. at the end) of the observation time. The conditions in Eq.
(\ref{eq:new_range_assumption}) reflect the assumption that ranges
at early (resp. late) time samples have non-negligible weights in
Eq. (\ref{eq:gamma11}) (resp. Eq. (\ref{eq:gamma22})) but they are
very similar to $r_{1}$ (resp. $r_{n}$), which is a reasonable assumption
in sonar tracking. Also, note that the mentioned condition is weaker
than $r_{i}\approx r$, $\forall i\in\mathcal{I}$, (i.e. an approximately
constant range assumption) and includes it as a more restrictive case.

\section{Choice of $\{\bm{\hat{i}}_{1},\bm{\hat{i}}_{n}\}$\label{sec:Appendix_choice-of_i0-in}}

The purpose of this Appendix is to show that: ($i$) $\bm{J}^{obs}$
contains only \emph{incomplete} information about $\{\bm{i}{}_{1},\bm{i}_{n}\}$,
meaning that it is not possible to extract these vectors without ambiguity;
($ii$) a good approach to extract such information is represented
by Eq. (\ref{eq:approx_i_i}). 

\begin{figure}
\centering{}\includegraphics[width=0.8\paperwidth]{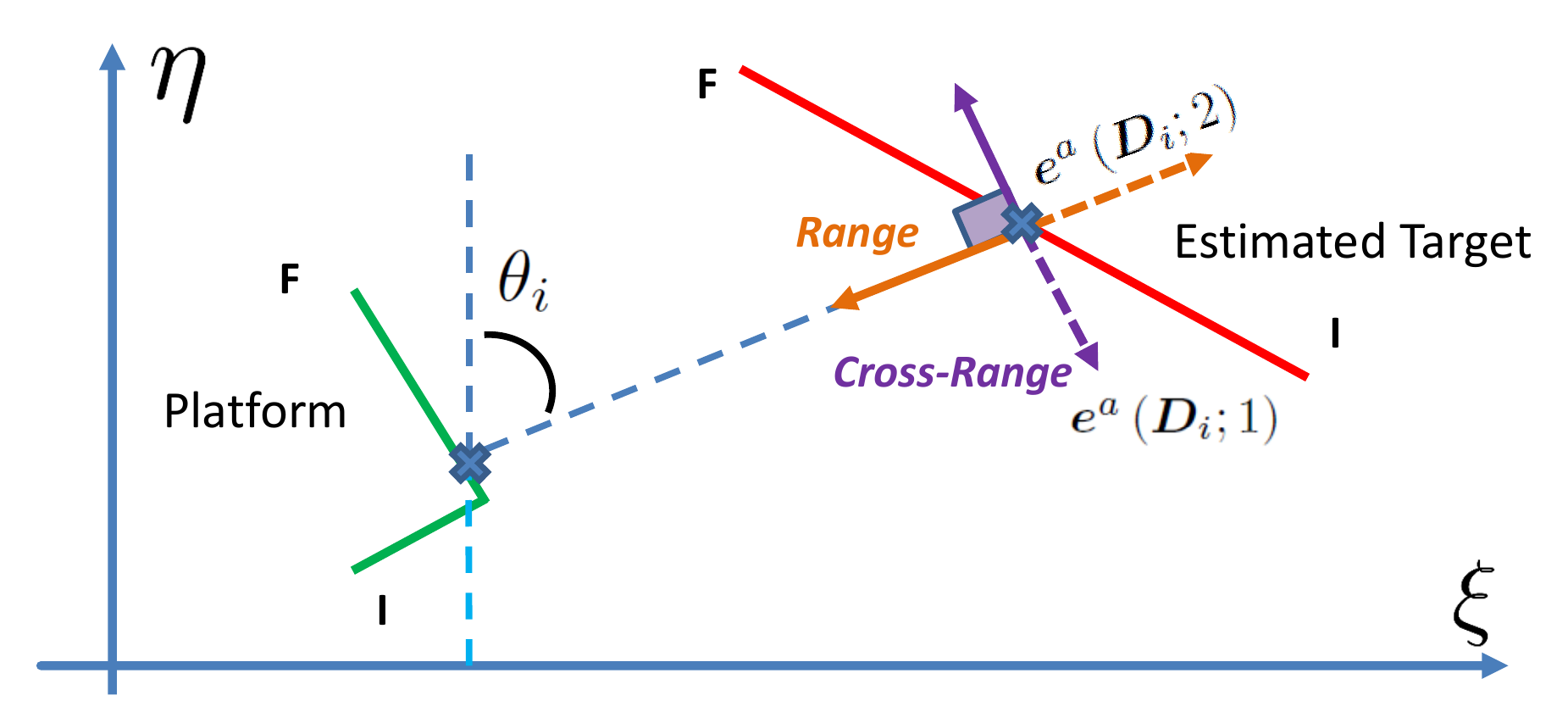}\caption{Eigenvectors ambiguity in the choice of $\{\bm{\hat{i}}_{1},\bm{\hat{i}}_{n}\}$.\label{fig:Eigenvectors-ambiguity}}
\end{figure}

For this purpose, let us consider the definition of $\bm{D}_{i}$
in Eqs. (\ref{eq:J11-block}) and (\ref{eq:J22-block}). It can be
readily shown that each $\bm{D}_{i}$ has eigenvalues $\{\lambda_{1,i},\lambda_{2,i}\}=\{1,0\}$
and that the corresponding (orthogonal) eigenvectors are
\begin{equation}
\bm{e}^{a}\left(\bm{D}_{i};1\right)=a\left[\begin{array}{c}
\cos\left(\theta_{i}\right)\\
-\sin\left(\theta_{i}\right)
\end{array}\right],\quad\bm{e}^{a}\left(\bm{D}_{i};2\right)=a\left[\begin{array}{c}
\sin\left(\theta_{i}\right)\\
\cos\left(\theta_{i}\right)
\end{array}\right],\qquad a\in\{-1,1\}\label{eq:eigenvectors_dyads}
\end{equation}
The pair $\left\{ \lambda_{1,i},\bm{e}^{a}\left(\bm{D}_{i};1\right)\right\} $
corresponds to the \emph{cross-range direction} at $t_{i}$, while
$\left\{ \lambda_{2,i},\bm{e}^{a}\left(\bm{D}_{i};2\right)\right\} $
corresponds to the \emph{range direction} at $t_{i}$. In fact, recall
that each $\bm{D}_{i}$ has an \emph{informative contribution} along
the \emph{cross-range} direction, thus $\lambda_{1,i}=1$%
\footnote{Note that even if the informative contribution of $\bm{D}_{i}$ is
always $\lambda_{1,i}=1$ along the cross-range direction, its contribution
in the FIM is weighted by $\frac{1}{r_{i}^{2}}$. %
}; on the other hand each $\bm{D}_{i}$ has\emph{ no informative contribution}
along the \emph{range} direction (since the range is estimated with
at least two bearing measurements), that is $\lambda_{2,i}=0$. For
this reason the information related to $\bm{i}_{1}$ (resp. $\bm{i}_{n}$)
is contained in $\bm{e}^{a}\left(\bm{D}_{1};2\right)$ (resp. $\bm{e}^{a}\left(\bm{D}_{n};2\right)$).
Note that however the sign-ambiguity in the definition of $\bm{e}^{a}\left(\bm{D}_{1};2\right)$
(resp. $\bm{e}^{a}\left(\bm{D}_{n};2\right)$) denotes the impossibility
of recovering exactly $\bm{i}_{1}$ (resp. $\bm{i}_{n}$), even if
$\bm{D}_{1}$ (resp. $\bm{D}_{n}$) had been perfectly available (a
graphical description is given in Fig. (\ref{fig:Eigenvectors-ambiguity}));
this proves ($i$).

Before proceeding in the proof, it is worth noting that $\bm{e}^{a}\left(b_{i}\bm{D}_{i};t\right)=\bm{e}^{a}\left(\bm{D}_{i};t\right)$,
$t\in\{1,2\}$, $b_{i}\in\mathbb{R}^{+}$; for such a reason in the
following w.l.o.g. we will search for a matrix $\bm{K}_{i}$ which
approximates well $\bm{D}_{i}$, except for a scale factor $b_{i}$,
i.e. $\bm{K}_{i}\approx b_{i}\bm{D}_{i}$; once $\bm{K}_{i}$ is obtained,
the pairs $\{\hat{\bm{i}}_{1}^{p},\hat{\bm{i}}_{n}^{q}\}$, $p,q\in\{-1,1\}$,
are simply evaluated by considering the least informative eigenvectors
of $\bm{K}_{1}$ and $\bm{K}_{n}$, respectively.

To obtain good estimates of $b_{1}\bm{D}_{1}$ and $b_{n}\bm{D}_{n}$
we can consider
\begin{eqnarray}
\bm{K}_{1} & \triangleq & \bm{J}^{obs}[1,1]=\frac{\breve{\alpha}_{\theta}}{r_{1}^{2}}\bm{D}_{1}+\breve{\alpha}_{\theta}\sum_{i=2}^{n-1}\frac{(1-\alpha_{i})^{2}}{r_{i}^{2}}\bm{D}_{i}\label{eq:scaled_approx_dyad_D0}\\
\bm{K}_{n} & \triangleq & \bm{J}^{obs}[2,2]=\frac{\breve{\alpha}_{\theta}}{r_{n}^{2}}\bm{D}_{n}+\breve{\alpha}_{\theta}\sum_{i=1}^{n-1}\frac{\alpha_{i}{}^{2}}{r_{i}^{2}}\bm{D}_{i}\label{eq:scaled_approx_dyad_Dn}
\end{eqnarray}
where in Eq. (\ref{eq:scaled_approx_dyad_D0}) (resp. Eq. (\ref{eq:scaled_approx_dyad_Dn}))
we stress the (scaled) contribution of $\bm{D}_{1}$ (resp. $\bm{D}_{n}$)
w.r.t. the spurious terms, i.e. $\bm{D}_{i}$, $i\in\mathcal{I}\backslash\{1\}$
(resp. $i\in\mathcal{I}\backslash\{n\}$). Exploiting again the inequalities
among $\alpha_{i}$ and the assumptions of Eq. (\ref{eq:new_range_assumption}),
it can be shown that $\frac{\breve{\alpha}_{\theta}}{r_{1}^{2}}\bm{D}_{1}$
(resp. $\frac{\breve{\alpha}_{\theta}}{r_{n}^{2}}\bm{D}_{n}$) is
well approximated by Eq. (\ref{eq:scaled_approx_dyad_D0}) (resp.
Eq. (\ref{eq:scaled_approx_dyad_Dn})); note that in this case convex
combination counterparts are not needed because eigenvectors are not
changed by a scaling factor. However, we will show hereinafter that
a better estimate of $b_{i}\bm{D}_{i}$ can be obtained.

In fact let us consider $\bm{C}^{obs}\triangleq\left(\bm{J}^{obs}\right){}^{-1}$
and denote $\bm{C}^{obs}[\ell,m]$ as the $(\ell,m)$th $[2\times2]$
block matrix of $\bm{C}^{obs}$. By exploiting the block-wise inversion
formula \cite{Bernstein2005} we obtain 
\begin{eqnarray}
(\bm{C}^{obs}[1,1])^{-1} & = & \bm{J}^{obs}[1,1]-\bm{J}^{obs}[1,2]\bm{J}^{obs}[2,2]^{-1}\bm{J}^{obs}[1,2]\\
(\bm{C}^{obs}[2,2])^{-1} & = & \bm{J}^{obs}[2,2]-\bm{J}^{obs}[1,2]\bm{J}^{obs}[1,1]^{-1}\bm{J}^{obs}[1,2]
\end{eqnarray}
Exploiting the expression for $\bm{J}[\ell,m]$, $\ell,m\in\{1,2\}$
as in Eqs. (\ref{eq:J11-block}) and (\ref{eq:J22-block}) and putting
in evidence the scaled contribution of $\bm{D}_{1}$ (resp. $\bm{D}_{n}$),
we get:
\begin{align}
(\bm{C}^{obs}[1,1])^{-1} & =\frac{\breve{\alpha}_{\theta}}{r_{1}^{2}}\bm{D}_{1}+\breve{\alpha}_{\theta}\left\{ \sum_{i=2}^{n-1}\frac{1}{r_{i}^{2}}\bm{D}_{i}\left[(1-\alpha_{i})^{2}\bm{I}_{2}-\alpha_{i}(1-\alpha_{i})\bm{T}_{1}\right]\right\} \label{eq:improved_scaled_approximation_dyad_D0}\\
(\bm{C}^{obs}[2,2])^{-1} & =\frac{\breve{\alpha}_{\theta}}{r_{n}^{2}}\bm{D}_{n}+\breve{\alpha}_{\theta}\left\{ \sum_{i=2}^{n-1}\frac{1}{r_{i}^{2}}\bm{D}_{i}\left[\alpha_{i}{}^{2}\bm{I}_{2}-\alpha_{i}(1-\alpha_{i})\bm{T}_{n}\right]\right\} \label{eq:improved_scaled_approximation_dyad_Dn}
\end{align}
where $\bm{T}_{1}$ and $\bm{T}_{n}$ are defined respectively as
\begin{eqnarray}
\bm{T}_{1} & \triangleq & \left[\left(\sum_{\ell=2}^{n}\frac{\alpha_{\ell}{}^{2}}{r_{\ell}^{2}}\bm{D}_{\ell}\right)^{-1}\sum_{j=2}^{n-1}\frac{\alpha_{j}(1-\alpha_{j})}{r_{j}^{2}}\bm{D}_{j}\right]\\
\bm{T}_{n} & \triangleq & \left[\left(\sum_{\ell=1}^{n-1}\frac{(1-\alpha_{\ell}){}^{2}}{r_{\ell}^{2}}\bm{D}_{\ell}\right)^{-1}\sum_{j=2}^{n-1}\frac{\alpha_{j}(1-\alpha_{j})}{r_{j}^{2}}\bm{D}_{j}\right]
\end{eqnarray}
It is apparent how each spurious term in the braces of Eqs. (\ref{eq:improved_scaled_approximation_dyad_D0})
and (\ref{eq:improved_scaled_approximation_dyad_Dn}) (cf. with Eqs.
(\ref{eq:scaled_approx_dyad_D0}) and (\ref{eq:scaled_approx_dyad_Dn}))
is now filtered through the matrix gain $\bm{T}_{1}$ (resp. $\bm{T}_{n}$)
with weight $\alpha_{i}(1-\alpha_{i})$. The matrix $\bm{T}_{1}$
(resp. $\bm{T}_{n}$) represents a ``smoothing'' factor (independent
of $t_{i}$, cf. with Eqs. (\ref{eq:scaled_approx_dyad_D0}) and (\ref{eq:scaled_approx_dyad_Dn})).
The weights $\alpha_{i}(1-\alpha_{i})$ are such that there is a higher
correction w.r.t. $\bm{D}_{i}$ corresponding to the the middle of
the observation interval, while a little correction at the beginning
or the end of the observation interval. However, while in the first
case the correction tends to reduce the spurious term, in the latter
case there is an increase of the error given by the spurious term.

To gain  intuition about the effect of matrix $\bm{T}_{1}$ (same
considerations apply to $\bm{T}_{n}$ in Eq. (\ref{eq:improved_scaled_approximation_dyad_Dn}))
on Eq. (\ref{eq:improved_scaled_approximation_dyad_D0}) let us consider
the case $r_{i}\approx r$, $\forall i\in\mathcal{I}$. In this case
we have that $\bm{T}_{1}\approx\left(\sum_{\ell=2}^{n}\alpha_{\ell}{}^{2}\bm{D}_{\ell}\right)^{-1}\sum_{j=2}^{n-1}\alpha_{j}(1-\alpha_{j})\bm{D}_{j}$
and the the sum of the spurious terms in Eq. (\ref{eq:improved_scaled_approximation_dyad_D0})
reduces to
\begin{equation}
\breve{\alpha}_{\theta}\left\{ \frac{1}{r^{2}}\sum_{i=2}^{n-1}\bm{D}_{i}\left[(1-\alpha_{i})^{2}\bm{I}_{2}-\alpha_{i}(1-\alpha_{i})\bm{T}_{1}\right]\right\} \label{eq:residual_term_dyads_dummy}
\end{equation}
Thus the magnitude of $\bm{T}_{1}$ will depend on the ratio of the
concentration of $t_{i}$ at the middle of the observation interval
by the concentration of $t_{i}$ at the end of the observation interval;
in fact a higher ratio will imply a lower distortion in the smoothing
of residual terms. 

\emph{Remark}: Note that Eqs. (\ref{eq:improved_scaled_approximation_dyad_D0})
and (\ref{eq:improved_scaled_approximation_dyad_Dn}) cannot be exploited
to obtain better estimates of $\hat{r}_{1}$ and $\hat{r}_{n}$ than
the ones in Eq. (\ref{eq:sec_approx_ri}), even if the spurious terms
are smoothed in such a case. The reason is that a proper normalization
factor to obtain a convex combination cannot be found, since it can
be shown that such a value would be dependent on $\bm{D}_{i}$, $i\in\mathcal{I}$,
which are clearly not available.

\section{Choice of $\bm{\hat{p}}_{p}(t_{k})$\label{sec:Appendix_choice of pptk}}

In order to obtain $\hat{\bm{p}}_{P}(t_{k})$ we will first seek an
approximation of $\bm{p}_{P}(t_{m})$ (denoted as $\bm{\tilde{p}}_{P}(t_{m})$),
where $t_{m}\triangleq\arg\min_{t_{i}\in\mathcal{T}}\left\Vert t_{i}-\frac{t_{n}-t_{1}}{2}\right\Vert _{2}$,
i.e. the nearest $t_{i}$ to the middle of the observation interval.
Similarly as $\{\bm{\breve{p}}_{P}(t_{1}),\bm{\breve{p}}_{P}(t_{n})\}$
(cf. Eq. (\ref{eq:ppt0_pptn_vector_range_form})), we can express
$\bm{\breve{p}}_{P}(t_{m})$ as 
\begin{equation}
\bm{\breve{p}}_{P}(t_{m})=\bm{\hat{p}}_{T}(t_{m})+r_{m}\bm{i}_{m}\label{eq:pp_tm_appendix}
\end{equation}
 The estimate of $r_{m}$, denoted as $\tilde{r}_{m}$, is obtained,
in analogy to Eq. (\ref{eq:sec_approx_ri}), as 
\begin{eqnarray}
\tilde{r}_{m} & \triangleq & \sqrt{\frac{\breve{\alpha}_{\theta}\sum_{i=1}^{n}\alpha_{i}(1-\alpha_{i})}{\mathrm{tr}\left(\bm{J}^{obs}[1,2]\right)}}\label{eq:range_estimate_pptk_bad}\\
\mathrm{tr}\left(\bm{J}^{obs}[1,2]\right) & = & \breve{\alpha}_{\theta}\left[\frac{\alpha_{m}(1-\alpha_{m})}{r_{m}^{2}}+\sum_{i=1,i\neq m}^{n}\frac{\alpha_{i}(1-\alpha_{i})}{r_{i}^{2}}\right]\label{eq:bad_range_estimate_tpoint}
\end{eqnarray}
As opposed to the case of $\{\bm{i}_{1},\bm{i}_{n}\}$, the estimate
of $\bm{i}_{m}$, denoted as $\bm{\tilde{i}}_{m}$, is obtained exploiting
$\bm{J}^{obs}[1,2]$ as a rough estimate of $b_{m}\bm{D}_{m}$, similarly
as in Eqs. (\ref{eq:scaled_approx_dyad_D0}) and (\ref{eq:scaled_approx_dyad_Dn}):
\begin{eqnarray}
\tilde{\bm{i}}_{m} & \triangleq & \bm{e}^{q}\left(\bm{J}^{obs}[1,2];2\right)\label{eq:direction_estimate_pptk_bad}\\
\bm{J}^{obs}[1,2] & = & \breve{\alpha}_{\theta}\left[\frac{\alpha_{m}(1-\alpha_{m})}{r_{m}^{2}}\bm{D}_{m}+\sum_{i=1,i\neq m}^{n}\frac{\alpha_{i}(1-\alpha_{i})}{r_{i}^{2}}\bm{D}_{i}\right]\label{eq:InverseFIMturningpoint}\\
q & \ni & \mathrm{sign}\left\langle \bm{e}^{j}\left(\bm{J}^{obs}[1,2];1\right),\bm{u}\right\rangle =\mathrm{sign}\left\langle \hat{\bm{i}}_{1},\bm{u}\right\rangle ,\quad j\in\{-1,1\}
\end{eqnarray}

The last line accounts for the sign ambiguity, in analogy to Eq. (\ref{eq:i_i_feasibility}).
The explicit form of $\bm{\tilde{p}}_{P}(t_{m})$ is obtained by replacing
$\{r_{m},\bm{i}_{m}\}$ with $\{\tilde{r}_{m},\bm{\tilde{i}}_{m}\}$
in Eq. (\ref{eq:pp_tm_appendix}). Some important remarks, about $\bm{\tilde{p}}_{P}(t_{m})$,
are in the following:
\begin{itemize}
\item Eq. (\ref{eq:bad_range_estimate_tpoint}) represents a weighted combination
of the inverse squared ranges, in analogy to Eq. (\ref{eq:sec_approx_ri}).
However it is apparent that the weight corresponding to $\frac{1}{r_{m}^{2}}$,
that is $\breve{\alpha}_{\theta}\alpha_{m}(1-\alpha_{m})$, is weaker
in this case w.r.t. the weights of the spurious terms. This leads
not only to a poorer estimate $\tilde{r}_{m}$, as opposed to $\hat{r}_{1}$
and $\hat{r}_{n}$, but also $\tilde{r}_{m}$ will be biased toward
the line between $r_{0}$ and $r_{n}$, leading to an estimated low-observable
platform trajectory (which is not desirable as an initial guess);
\item In Eq. (\ref{eq:InverseFIMturningpoint}) $\bm{J}^{obs}[1,2]$ is
used as a $b_{m}\bm{D}_{m}$ estimate, since a better estimate cannot
be found, differently from $\{\bm{D}_{1},\bm{D}_{n}\}$. In fact it
can be shown that $(\bm{C}^{obs}[1,2])^{-1}$ does not represent a
better estimate of $b_{m}\bm{D}_{m}$. Therefore $\bm{J}^{obs}[1,2]$
represents the only existing approximation of $b_{m}\bm{D}_{m}$. 
\end{itemize}
The above considerations suggest that one select a reasonable estimate
of $\bm{\breve{p}}_{P}(t_{k})$ according to a different approach,
as described in the following. 

In fact, we assume that the two legs form a $\pm\frac{\pi}{2}$ angle,
since the platform needs to perform a maneuver that guarantees a good
degree of observability; therefore $\{\bm{\breve{p}}_{P}(t_{1}),\bm{\breve{p}}_{P}(t_{k}),\bm{\breve{p}}_{P}(t_{n})\}$
will form a right triangle. Furthermore the sign ambiguity in the
turn leads to the definition of two specular vectors, denoted as $\bm{\rho}_{\ell}(t_{k})$,
$\ell\in\{-1,1\}$. 

Since $t_{k}$ is assumed known and $s$ is constant during the two
legs, it can be shown, after geometric considerations, that such vectors
are given by:
\begin{align}
\bm{\rho}_{\ell}(t_{k}) & \triangleq\hat{\bm{p}}_{P}(t_{1})+\frac{(t_{k}-t_{1})}{(t_{n}-t_{k})}\left\Vert \hat{\bm{p}}_{P}(t_{n})-\hat{\bm{p}}_{P}(t_{1})\right\Vert _{2}\cos\left(\nu\right)\left[\begin{array}{c}
\sin\left(\psi_{\ell}\right)\\
\cos\left(\psi_{\ell}\right)
\end{array}\right]\label{eq:ambiguous_guess_pptk-1}\\
\psi_{\ell} & \triangleq\arctan_{2}\left(\hat{\bm{p}}_{P}(t_{n})-\hat{\bm{p}}_{P}(t_{1})\right)+\ell\cdot\left(\frac{\pi}{2}-\nu\right)\label{eq:direction_ambiguous_pptk-1}\\
\nu & \triangleq\arctan\left(\frac{t_{k}-t_{1}}{t_{n}-t_{k}}\right)\label{eq:beta_angle-1}
\end{align}
where $\nu$ and $\frac{(t_{k}-t_{1})}{(t_{n}-t_{k})}\left\Vert \hat{\bm{p}}_{P}(t_{n})-\hat{\bm{p}}_{P}(t_{1})\right\Vert _{2}\cos\left(\nu\right)$
represent the angle whose vertex is $\hat{\bm{p}}_{P}(t_{n})$ and
the distance between $\hat{\bm{p}}_{P}(t_{1})$ and $\bm{\rho}_{\ell}(t_{k})$,
respectively. Finally, the unit vector $\left[\begin{array}{cc}
\sin\left(\psi_{\ell}\right) & \cos\left(\psi_{\ell}\right)\end{array}\right]^{t}$ represents the direction from $\hat{\bm{p}}_{P}(t_{1})$ to $\bm{\rho}_{\ell}(t_{k})$
and accounts for the sign ambiguity in the turn, through the angle
$\psi_{\ell}$.

The ambiguity is resolved by exploiting the coarse information given
by Eqs. (\ref{eq:bad_range_estimate_tpoint}) and (\ref{eq:InverseFIMturningpoint}):
\begin{eqnarray}
\hat{\bm{p}}_{P}(t_{k}) & = & \bm{\rho}_{q}(t_{m})\label{eq:final_turning_position_vector-1}\\
q & = & \arg\min_{\ell\in\{-1,1\}}\left\Vert \tilde{\bm{p}}{}_{P}(t_{m})-\bm{\rho}_{\ell}(t_{m})\right\Vert _{2}
\end{eqnarray}
where $\bm{\rho}_{\ell}(t_{m})$ denotes the position vector at $t_{m}$
of the two-leg trajectory described by\\
 $\{\hat{\bm{p}}_{P}(t_{1}),\bm{\rho}_{\ell}(t_{k}),\hat{\bm{p}}_{P}(t_{n})\}$.

\bibliographystyle{IEEEtranS}
\bibliography{IEEEabrv,MLPDAlocalizationsonar}

\end{document}